\documentclass[aps,pre,twocolumn]{revtex4} 

\usepackage{graphicx}
\usepackage{amsmath}
\usepackage{amsthm} 

\usepackage[utf8]{inputenc}
\usepackage{amssymb}

\usepackage{mathtools}

\usepackage{natbib}

\begin{document}

\title{Whistler wave generation by non$-$gyrotropic, relativistic, electron beams}

\author{M. Skender and D. Tsiklauri}
\affiliation{School of Physics and Astronomy, Queen Mary University of London,
327 Mile End Road, London, E1 4NS, United Kingdom}

\begin{abstract} 

Particle--in--cell code, EPOCH, is used for studying features of the wave component evident to propagate
backwards from the front of the non--gyrotropic, relativistic beam of electrons injected in the Maxwellian,
magnetised background plasma with decreasing density profile. 
According to recent findings presented in Tsiklauri (2011), Schmitz $\&$ Tsiklauri (2013) 
and Pechhacker $\&$ Tsiklauri (2012), in a 1.5--dimensional magnetised plasma system, 
the non--gyrotropic beam generates freely escaping  
electromagnetic radiation with properties similar to the Type--III solar radio bursts. 
In this study the backwards propagating wave component evident in the perpendicular components 
of the elecromagnetic field in such a system is presented for the first time.   
Background magnetic field strength in the system is varied in order to prove that the backwards propagating 
wave's frequency,
prescribed by the whistler wave dispersion relation,  
is proportional to the specified magnetic field. 
Moreover, the identified whistlers are shown to be generated by the
normal Doppler-shifted relativistic resonance. Large fraction of the energy of the perpendicular 
electromagnetic field components is found to be carried away by the whistler waves, while a small but sufficient
fraction is going into L-- and R-- electromagnetic modes.

\end{abstract}

\maketitle

\section{Introduction} \label{SecIntroduction}

Narow--band radio emissions originating from the Sun with well expressed impulsive onset at each frequency, 
known as the Type--III bursts, are characterised by well defined frequency drift and sometimes a separation 
of the emission pattern into two harmonics
\cite{1979itsr.book.....K}.
Super$-$thermal electron beams travelling away from the Sun on the open magnetic field lines
 are widely accepted to be the source of the Type--III bursts \cite{1981ApJ...251..364L,1984A&A...141...30D}.
The first and the most widely accepted mechanism for the generation
of the Type--III bursts was based on the plasma emission. A fast moving ($0.2c - 0.5c$)
electron beam excites Langmuir waves at the local
plasma frequency, $\omega_{p}$. The Langmuir waves are partially transformed 
via scattering at $\omega_{p}$ and $2 \omega_{p}$,
with ion sound and oppositely propagating Langmuir waves, respectively, 
into electromagnetic waves. As the electron beam propagates away from the Sun, 
through less dense coronal and interplanetary environment, the frequency of the emitted electromagnetic radiation decreases, because plasma frequency is a function of the square root of the plasma density.

Type$-$III bursts have been subject of theoretical, observational and numerical studies \cite{1987SoPh..111...89M,
2002SSRv..101....1A,2008SoPh..253....3N,2008A&ARv..16....1P}.
The first detailed theory of the Type$-$III emission invoked coherent plasma waves, generated by a stream 
of fast particles \cite{1958AZh....35..694G}, which are  
due to Rayleigh and combination scattering at $\omega_{p}$ and $2 \omega_{p}$  
subsequently transformed into radio waves. Stochastic growth of the density irregularities
was invoked in order to produce stochastically generated clumpy Langmuir waves, where the ambient density
perturbations cause the beam to fluctuate around marginal stability \cite{1992SoPh..139..147R}.
Other mechanisms which generate the Type$-$III emission include: linear mode conversion of Langmuir
waves \cite{2005PhPl...12e2315C}, the antenna mechanism \cite{2010JGRA..115.1101M}
and non$-$gyroptropic electron beam emission \cite{2011PhPl...18e2903T}.

In Ref. \cite{2013PhPl...Schmitz} it was found that the non$-$gyrotropic beam excites electromagnetic radiation
by $(\omega, k)$$-$space drift (or in other words wave refraction)
while propagating along the 1$-$dimensional spatial domain throughout the decreasing plasma density profile.
The analysis presented here
reveals a new signal evident in the perpendicular components of the electromagnetic field in the scenario 
of the non$-$gyrotropic emission: A wave propagating backwards from the beam front.  
Investigation of the features of this wave component, its characterisation and its generation mechanism is topic
of the presented numerical study. The background magnetic field was varied in order to prove the
backwards travelling wave component's frequency to be proportional to the imposed background magnetic field.
Analysis of the $(\omega, k)$$-$space drift shows that the  wave fits well to the whistler dispersion curve.
The mechanism which generates the backwards propagating whistler waves is found to correspond to 
the normal Doppler-shifted relativistic resonance.
Energy partition indicates that most of the energy in the perpendicular
electromagnetic components is taken by the whistlers, while small but sufficient fraction is going
into L-- and R-- electromagnetic modes.

Whistlers are ubiquitous very low frequency (VLF) waves, detected in the Earth radiation belts,  
in some cases produced in the atmosphere by  lightning, studied
numerically and in the space and laboratory experiments 
\cite{1998PhPl....5.4243K,1997PhPl....4.4126V,2011JGRA..116.6306H}. 
Electron beams injected from satellites
into the ionospheric and magnetospheric plasma have resulted in emission of the whistlers mainly from
the front of the beams \cite{1999PhPl....Starodubtsev}. 
Emission of the whistlers through normal Doppler-shifted resonance has been reported
in a laboratory experiment, where
the excited whistlers propagate opposite to the beam direction and their phase 
and group velocities are characteristic of the beam$-$whistler resonant cyclotron coupling \cite{1999PhRvL..83.1335S}.   

The numerical model is described in Sec. II. In Sec. III the general evolution of the system 
and the evidence for the backwards propagating waves are presented, followed by their characterisation 
and identification through altering the background magnetic field.  
Further, the velocity and the  width of the injected beam are varied in order to prove
that our Doppler relativistic resonance excitation mechanism for whistler waves is valid.
Lastly, the partition of the total energy into the electromagnetic energy and sub--partition 
of the electromagnetic energy into whistlers and L-- and R-- electromagnetic modes is evaluated.
Results are summarised in Sec. IV.

\section{Simulation Setup \label{SecSimulation}}

Following recent works \cite{2011PhPl...18e2903T,2012PhPl...Pechhacker},
a 1.5--dimensional system of non--gyrotropic, super--thermal, electron  beam  injected into background
Maxwellian plasma is employed for studying wave generation in the context of the Type--III emission,
specifically to establish the $(\omega, k)$$-$space drift (wave refraction)  of transverse electromagnetic 
fields, caused by the density gradient. Fully
electromagnetic relativistic  particle--in--cell (PIC) code EPOCH with implemented open boundary conditions
 is utilised as in the study \cite{2013PhPl...Schmitz}. The code is developed by the collaborative
 computational plasma physics (CCPP)
consortium of the UK researchers, funded by the
Engineering and Physical Sciences Research Council (EPSRC).

The open magnetic field line between the Sun and the Earth, along which the super--thermal electron beams
generating the Type--III radio bursts propagate, is represented by a constant background magnetic field
$B_{0x}$. In the reference run the imposed value of the background magnetic field is 
 $B_{0x} = 3.208 \cdot 10^{-4} T$, as in \cite{2013PhPl...Schmitz}. 
We allow for the $x$$-$coordinate variation of the velocity components, $v_{x}$, $v_{y}$, $v_{z}$, 
of all particles as well as of the electromagnetic field components,
$E_{x}$, $E_{y}$, $E_{z}$, $B_{x}$, $B_{y}$ and $B_{z}$,
whereby the $y$-- and $z$-- coordinates are ignorable. 

The background plasma density close to the Sun is assumed to be $n_{Sun}=10^{14}$ $m^{-3}$ and at the
position of the Earth it drops to the $n_{Earth}=10^{6}$ $m^{-3}$, following the
parabolic profile:
\begin{equation}
n_{e}(x) = n_{Earth} + (n_{Sun} - n_{Earth}) \left( \frac{x}{L} - 1 \right)^2   \\,
\end{equation}
where $L$ stands for the total size of the simulation box and $x$ is spatial coordinate along 
the simulation domain. As a result of the above defined parameters, plasma beta in $x = 0$ is set to
$\beta = 0.01012$.

The beam is set as a separate electron population at the left side of the domain, representing the
side close to the Sun, with the density profile defined by:
\begin{equation}
n_{beam} =  n_{0}  e^{- \left( (x - L/25) / (L/40)\right)^8}   \\,
\end{equation}
where $n_{0}=10^{11}$ $m^{-3}$ is the beam density at $x = 0$. The beam density has 
the length of $L_{beam} = L/20$. The beam is injected at the beginning, at $t = 0$, and
propagates along the density gradient in the simulation  domain with a constant velocity.
The beam is not replenished, $i.e.$ we solve the initial value problem.
Spiky density profile of the beam at the end of the
simulation, as seen in Fig.~\ref{beam_rr}, is consistent with the quasi--linear theory prediction.
In the reference run the average speed of the electrons in 
the beam is taken to be
$v_{beam} = c/2 $, where $c$ is the speed of the light. The drift velocity is assumed to be equal in the 
perpendicular and $x-$ directions of the simulation domain, therefore the pitch angle $\vartheta_{p} 
= 45^{\circ}$. 
The temperature of the electron beam is taken to be $T_{beam} = 6 \cdot 10^6 K$ and the temperature of the
background plasma is $T_{e} = T_{p} = 3 \cdot 10^5 K$.

In the present study the three particle species, beam electrons, background electrons and background ions, are
 initialised each with 500 particles per cell.
The size of the physical box is $L_{x} = 245.69 m$ and the numerical box has $65000$ grid points. 
The grid size equals to the electron Debye length
found near the Sun at the location of the highest density, 
$\lambda_{De} = 0.00378 m$.

\section{Results}

\subsection{Backwards propagating wave}

\begin{figure}
\begin{center}
\includegraphics[width=7cm]{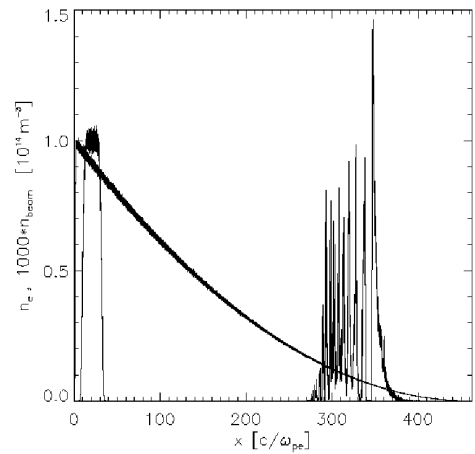} \\
\end{center}
\caption{Beam density multiplied by the $10^{3}$ at the beginning (left) and at the end (right) 
of the reference run 
as a function of distance $x$ normalised by $c/\omega_{pe}($x=0$)$ depicted by two towery profiles. 
The parabolic shaped curve
denotes the background plasma density profile. 
}
\label{beam_rr}
\end{figure}

In the reference run the choice of the background magnetic field $B_{0x} = 3.208 \cdot 10^{-4} T$ sets 
the ratio of the electron 
cyclotron and electron plasma frequency at the left side of the simulation domain 
to be $\omega_{ce} / \omega_{pe}($x=0$) = 0.1$. 
The numerically studied plasma is weakly magnetised.
The initialised beam traverses $\sim 3/4$ of the simulation domain in the course of the
simulation, as shown in Fig.~\ref{beam_rr}.

\begin{figure}
\begin{center}
\includegraphics[width=7cm]{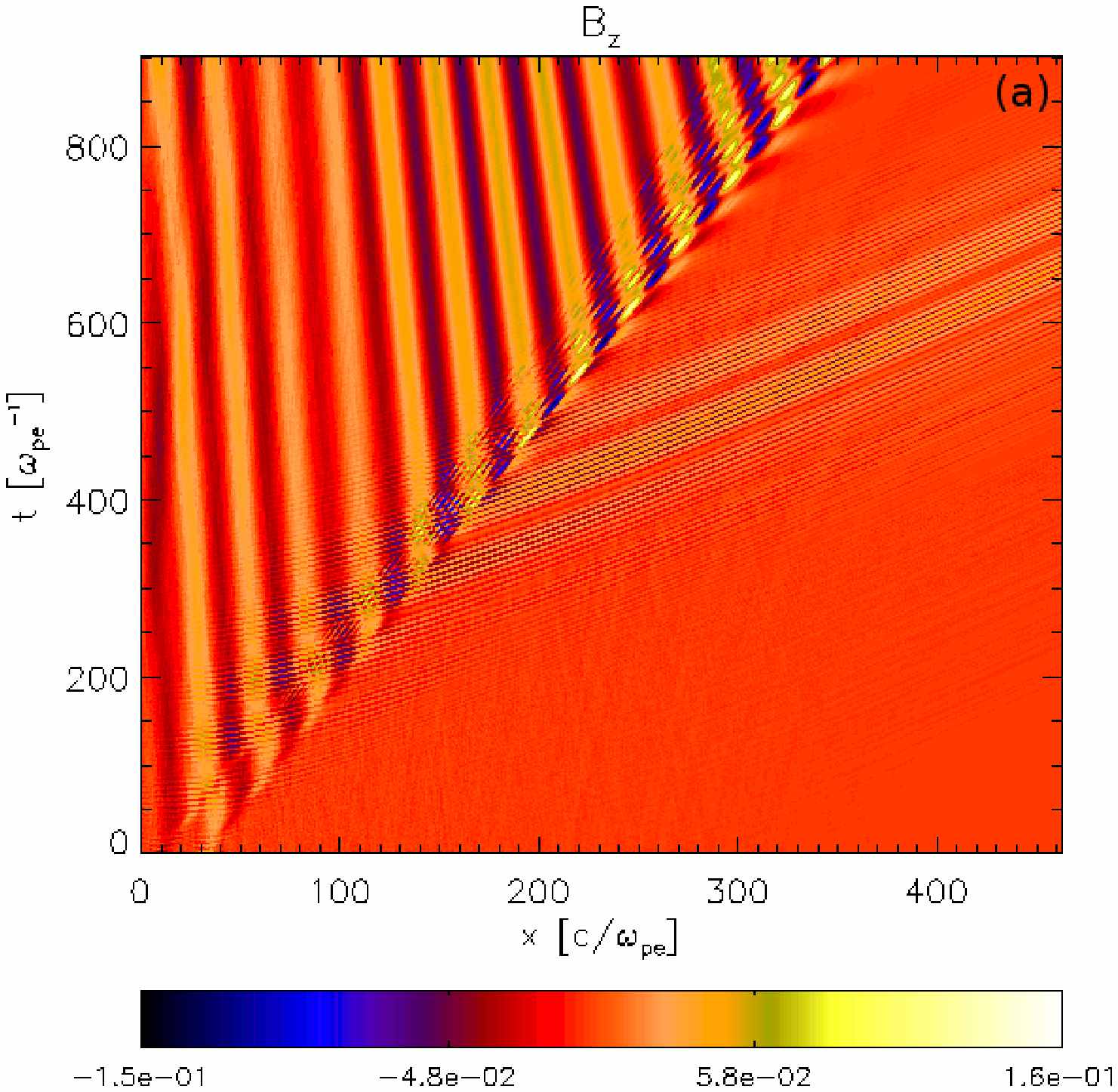} \\
\includegraphics[width=7cm]{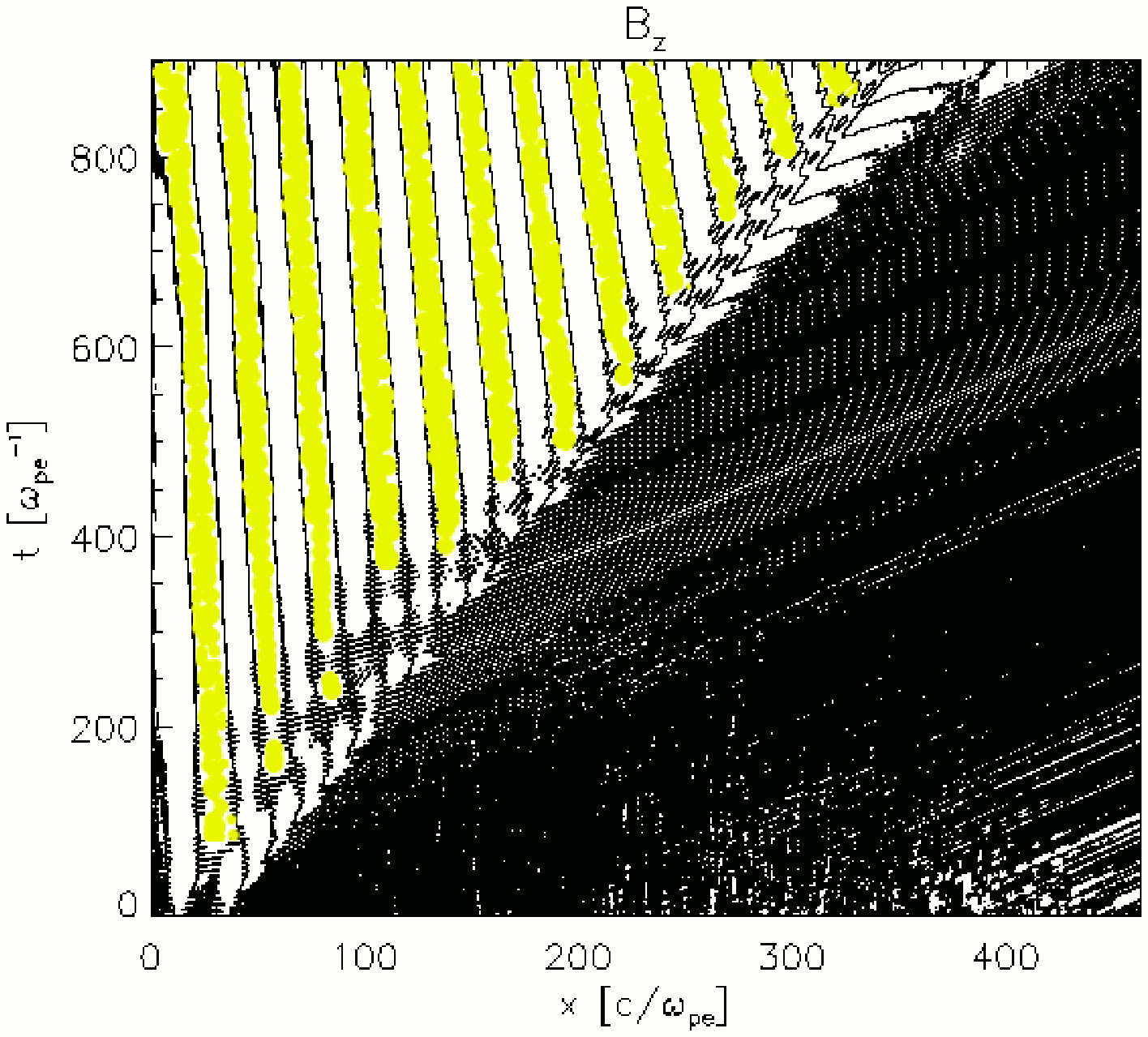} \\
\includegraphics[width=7cm]{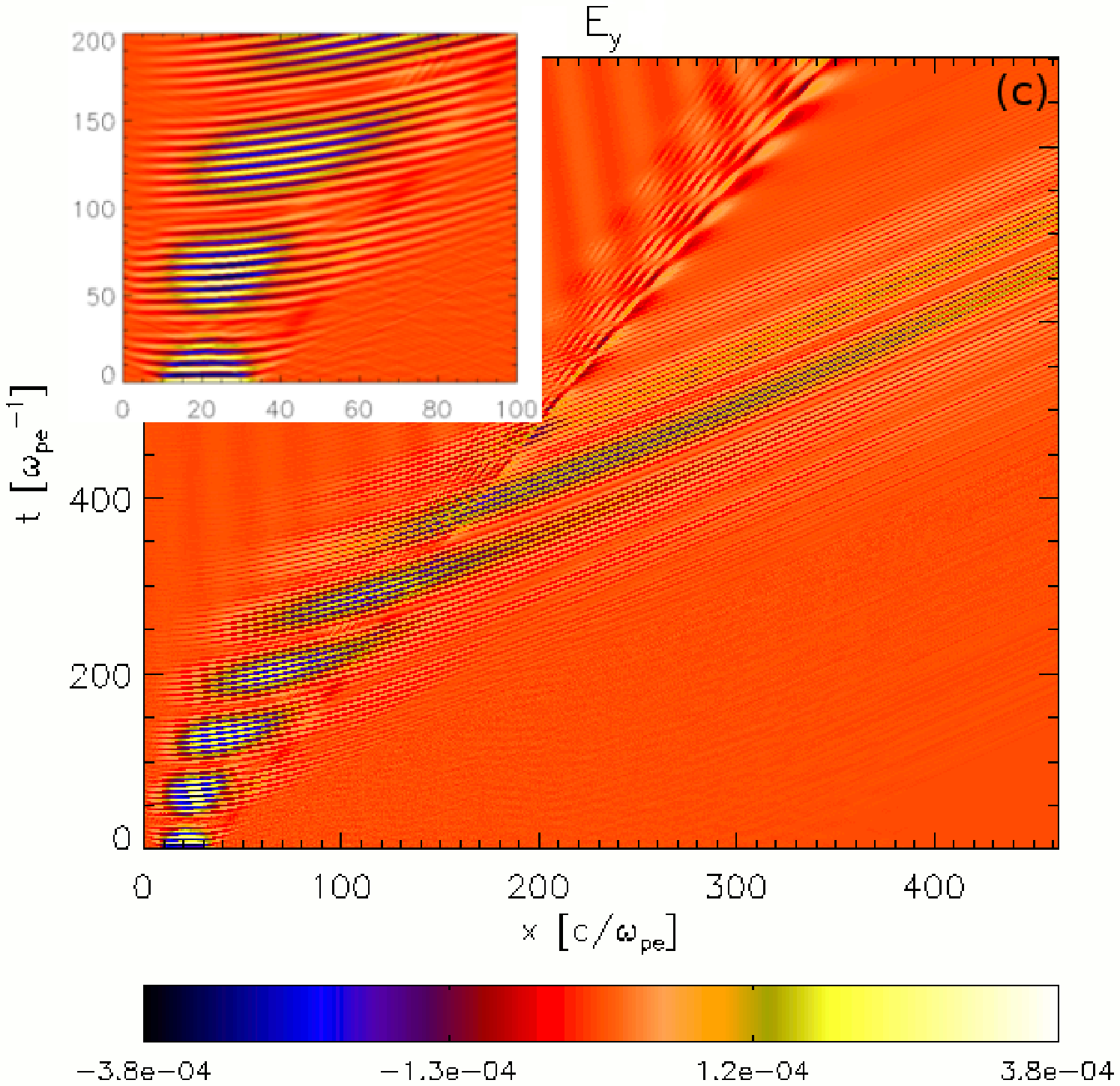} \\
\end{center}
\caption{(a) The magnetic field component $B_z$ as a function of space and time in
the reference run, where $\omega_{ce}/\omega_{pe}($x=0$) = 0.1$. 
Existence of the backwards travelling wave in upper left region is evident. 
(b) Contour plot of the time--distance plot for $B_z$, used for 
estimating the wavelength $\lambda$ of the backwards travelling wave. 
Beats are produced in the interaction of the $v_{beam,\perp}$ with the backwards propagating wave.
(c) Time$-$distance plot of $E_y$ for the same run.
The insert depicts magnified part of the injection region where the beats are formed. 
The escaping electromagnetic radiation evolved from the formed beat
of superposed $\omega_{R}(k_{beam})$ and $\omega_{L}(k_{beam})$. 
}
\label{Bz_rr}
\end{figure}

Wave activity present in the system of non--gyrotropic beam injected into the Maxwellian plasma 
with decreasing density profile  
evidenced by the perpendicular component of the magnetic field $B_z$  is presented in 
Figs.~\ref{Bz_rr} (a) and (b).
$B_z$, normalised by $e/m_{e}\omega_{pe}($x=0$)$, is plotted as a function of the
position $x$, normalised by $c/\omega_{pe}($x=0$)$, and the time $t$, normalised by $1/\omega_{pe}($x=0$)$. 
In Fig.~\ref{Bz_rr}~(a)  the beam is seen to propagate 
along the x$-$axis with the velocity $v_{beam,\parallel}=c/(2\sqrt{2})$.  
In the right half of the numerical box freely escaping electromagnetic waves with slope $v = c$
 are apparent after the time
$\approx$ 500 $\omega_{pe}(x=0)$, as in Refs.~\cite{2011PhPl...18e2903T},\cite{2013PhPl...Schmitz}. 

The wave found in the left hand side of the box is created at the front edge of the beam and
propagates $\it{backwards}$. Its wavelength $\lambda \approx$ 26.6$-$28.6 $[c/\omega_{pe}(x=0)]$
is inferred from the contour version  of the  $B_z$
time--distance plot, presented in Fig.~\ref{Bz_rr}~(b).
Fig.\ref{Bz_rr}~(c) presents the 
time-distance plot of perpendicular electric field component $E_y$, normalised by $e/cm_{e}\omega_{pe}($x=0$)$,
in which formation of the freely escaping electromagnetic wave is most clearly visible. Time-distance plots of 
perpendicular magnetic and electric field components exhibit the same wave activity pattern. However, 
backwards propagating wave is more pronounced feature in the perpendicular magnetic field, while 
the electromagnetic wave is more clearly seen from the perpendicular electric field. 

\begin{figure}
\begin{center}
\includegraphics[width=7cm]{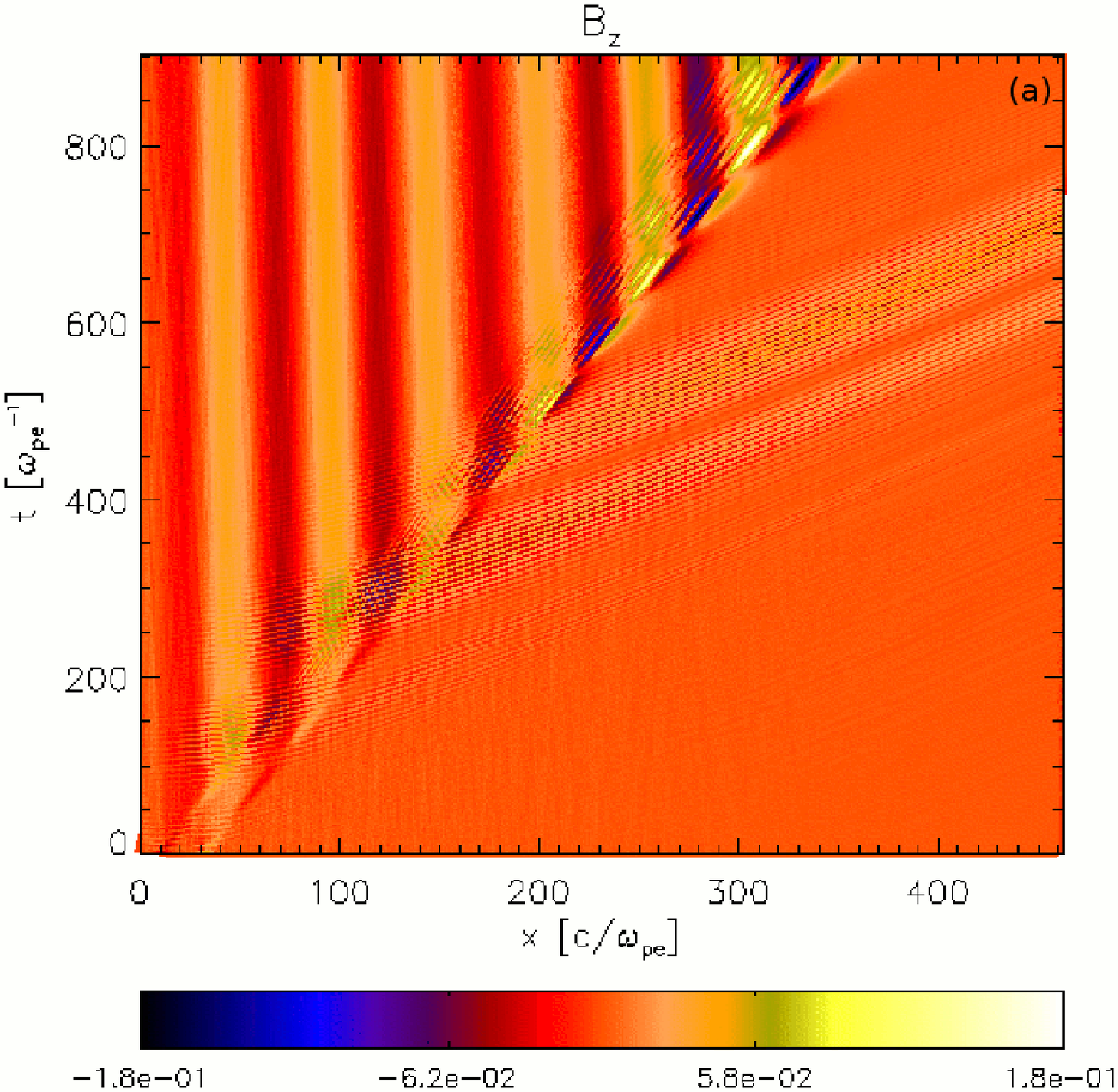} \\
\includegraphics[width=7cm]{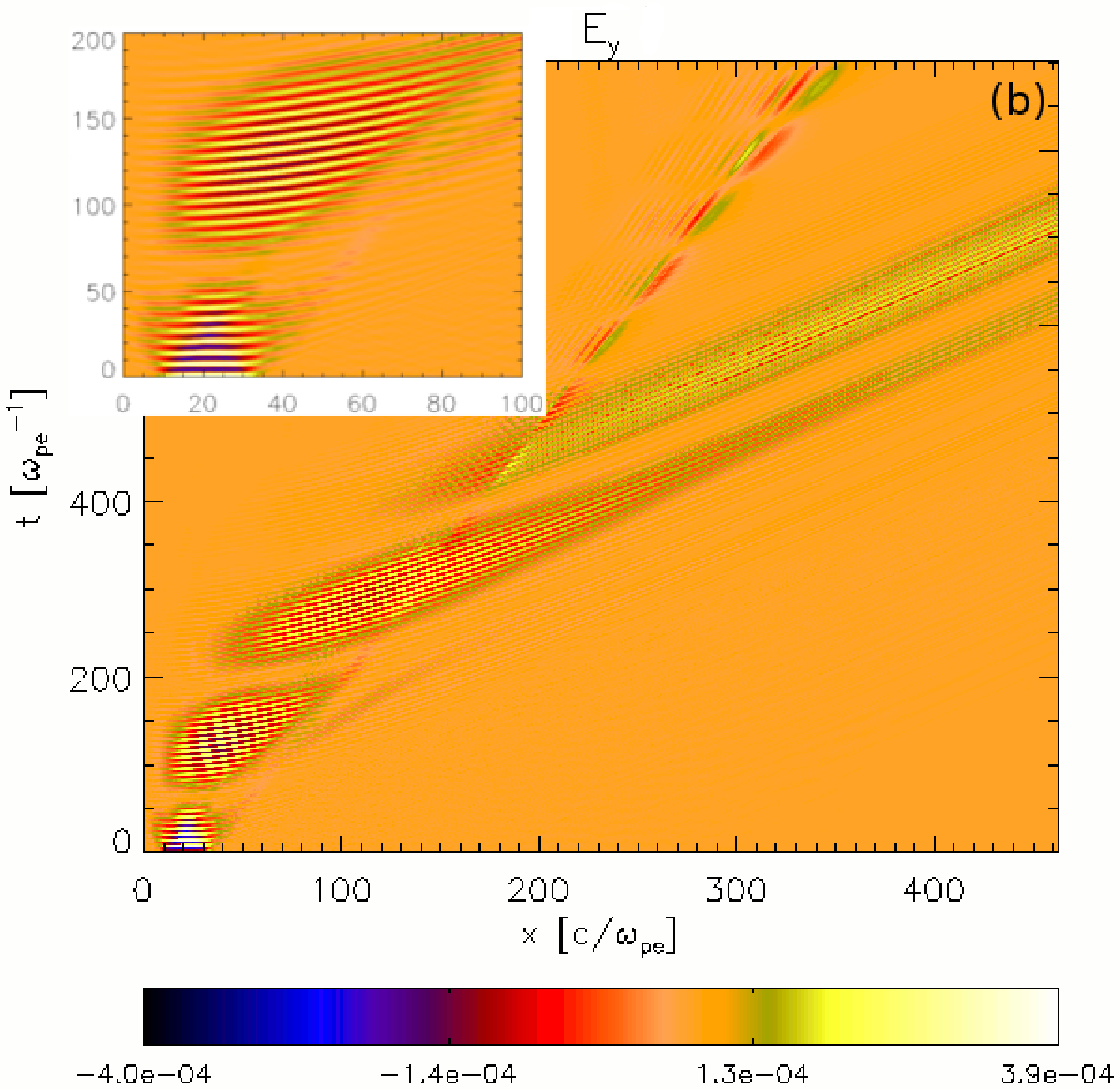} \\
\end{center}
\caption{(a) Time$-$distance plot for $B_z$ and  the $\omega_{ce}/\omega_{pe}($x=0$) = 0.05$.
Backwards propagating wave is seen to have $\approx$ 2 times larger wavelength than in
the reference run.
(b) Time$-$distance plot of $E_y$ for the same run.
The beats are formed by superposition of $\omega_L(k_{beam})$ and $\omega_R(k_{beam})$. 
}
\label{td_0_05}
\end{figure}

\begin{figure}
\begin{center}
\includegraphics[width=7cm]{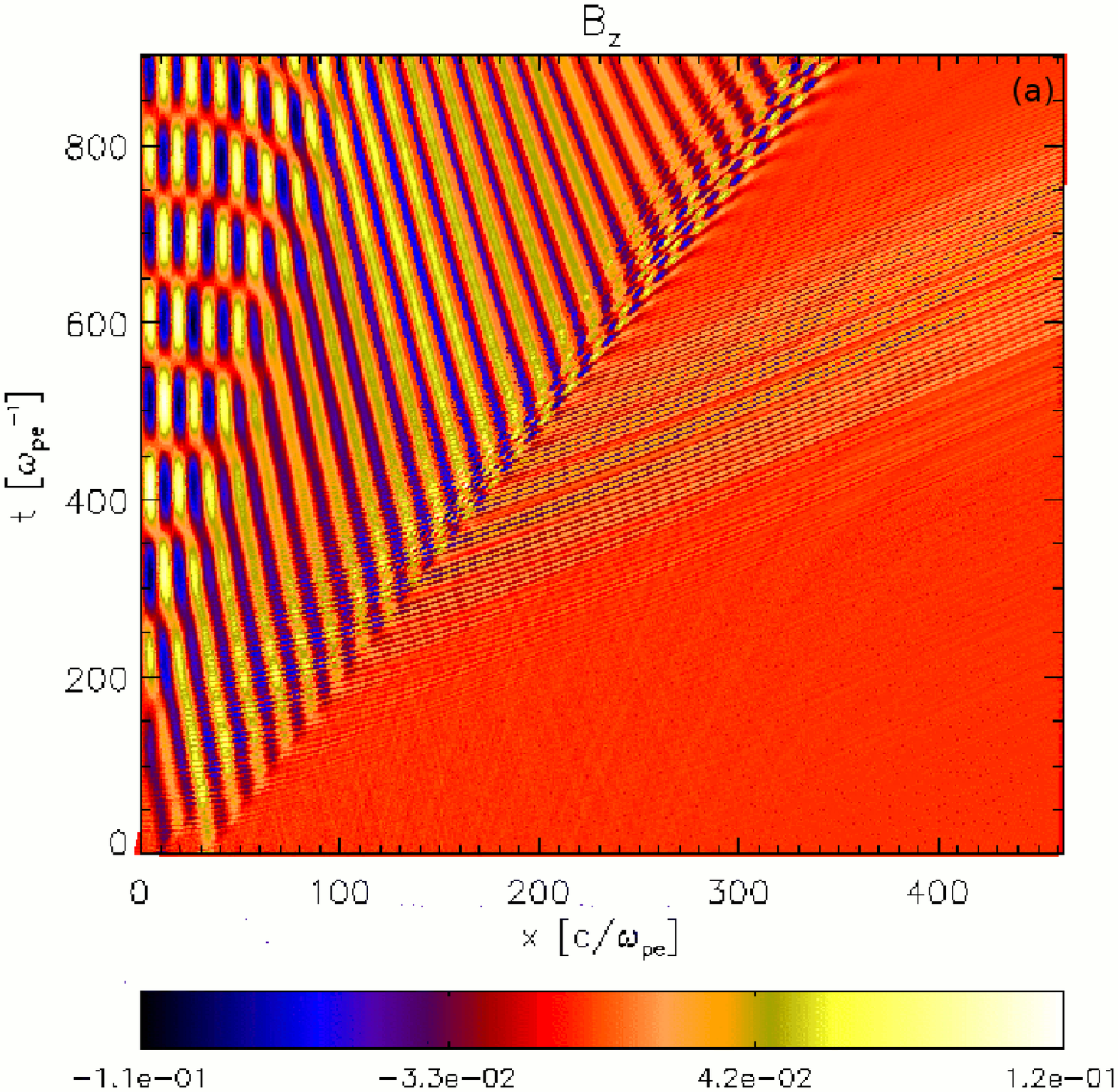} \\
\includegraphics[width=7cm]{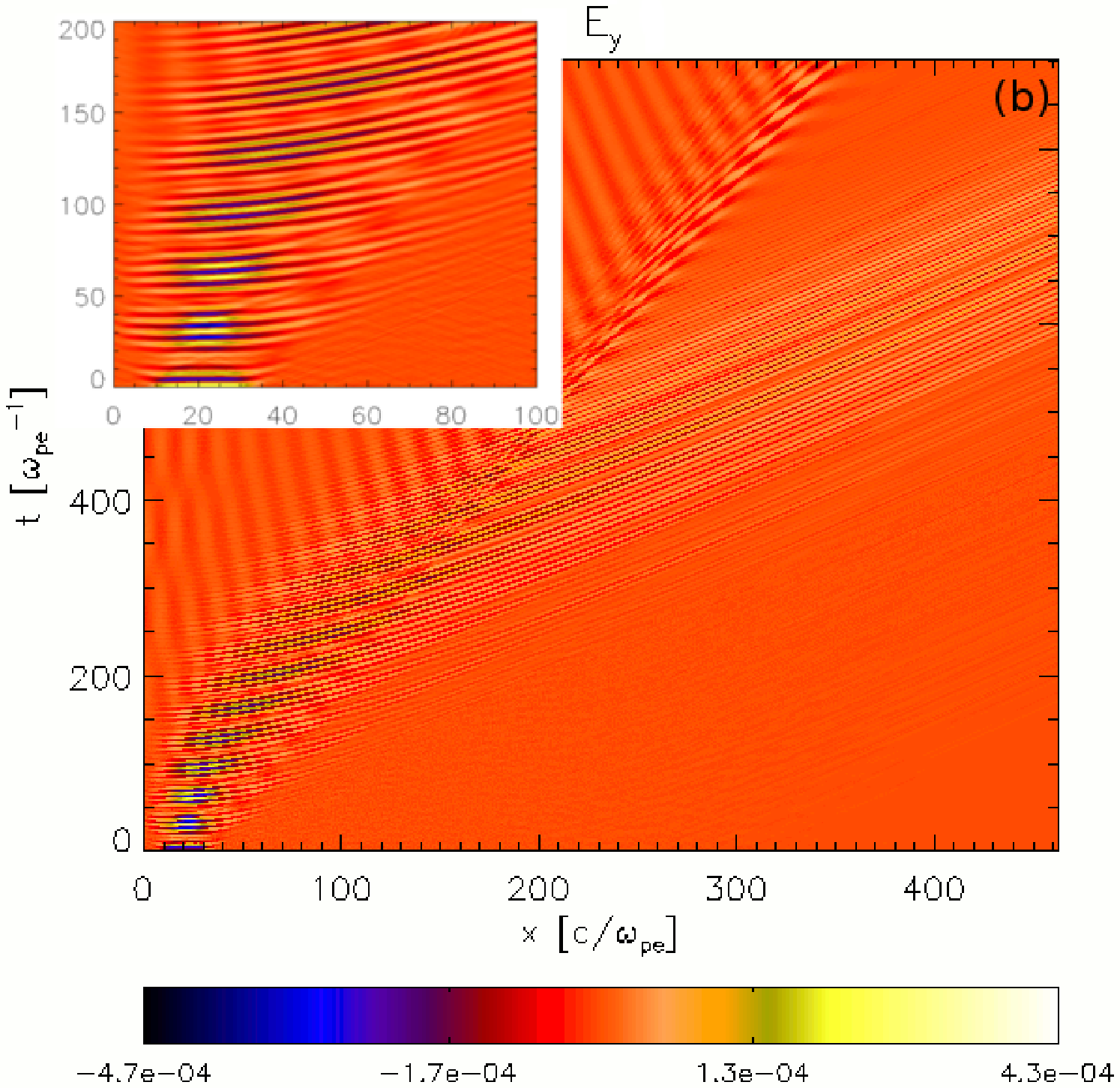} \\
\end{center}
\caption{(a) Time$-$distance plot for $B_z$ and  the $\omega_{ce}/\omega_{pe}($x=0$) = 0.2$.
Backwards propagating wave has wavelength $\approx$ 2 times smaller then the $\lambda$ in 
the reference run.
Reflection from the boundary is taking place at the left side of the domain.
(b) Time$-$distance plot of $E_y$ for the same run.
The beats are formed by superposition of $\omega_L(k_{beam})$ and $\omega_R(k_{beam})$. 
}
\label{td_0_2}
\end{figure}

In order to alter the properties of the generated backwards propagating wave, 
the background magnetic field $B_{0x}$ is altered.
Two times smaller $B_{0x}$ yields $\omega_{ce}/\omega_{pe}($x=0$) = 0.05$, while two times larger
$B_{0x}$ sets $\omega_{ce}/\omega_{pe}($x=0$) = 0.2$.  Time$-$distance plots of $B_z$ and $E_y$  
are presented respectively in  (a) and (b) panels of Figs.~\ref{td_0_05} and \ref{td_0_2}. 
Figs.~\ref{td_0_05}~(b) and \ref{td_0_2}~(b) depict formation of the wave beats
in the injection region.
Fig.~\ref{td_0_05}~(a) displays the case of the 2 times smaller
applied $B_{0x}$ and hence 2 times smaller $\omega_{ce}$. The $\lambda$ in this case is estimated to be
51.9$-$55.2 $[c/\omega_{pe}(x=0)]$. 
Fig.~\ref{td_0_2}~(a) contains the case with the $B_{0x}$ and $\omega_{ce}$ two times  
larger than in the reference run. The $\lambda$ is here $\approx$ 14.2$-$15.8 $[c/\omega_{pe}(x=0)]$.
When $\omega_{ce}$ is two times smaller than in the reference run, $\lambda$ is approximately two times
larger than inferred in the reference run. On the other hand, $\omega_{ce}$ two times larger yields 
$\lambda$ approximately two times smaller than in the reference run. 
Backwards propagating waves are thereby confirmed to have frequency proportional to the 
electron-cyclotron frequency $\omega_{ce}$, as prescribed by the dispersion relation (see 
Eq.~\ref{whistler_dispersion} below for details).

\begin{figure}
\begin{center}
\includegraphics[width=6.2cm]{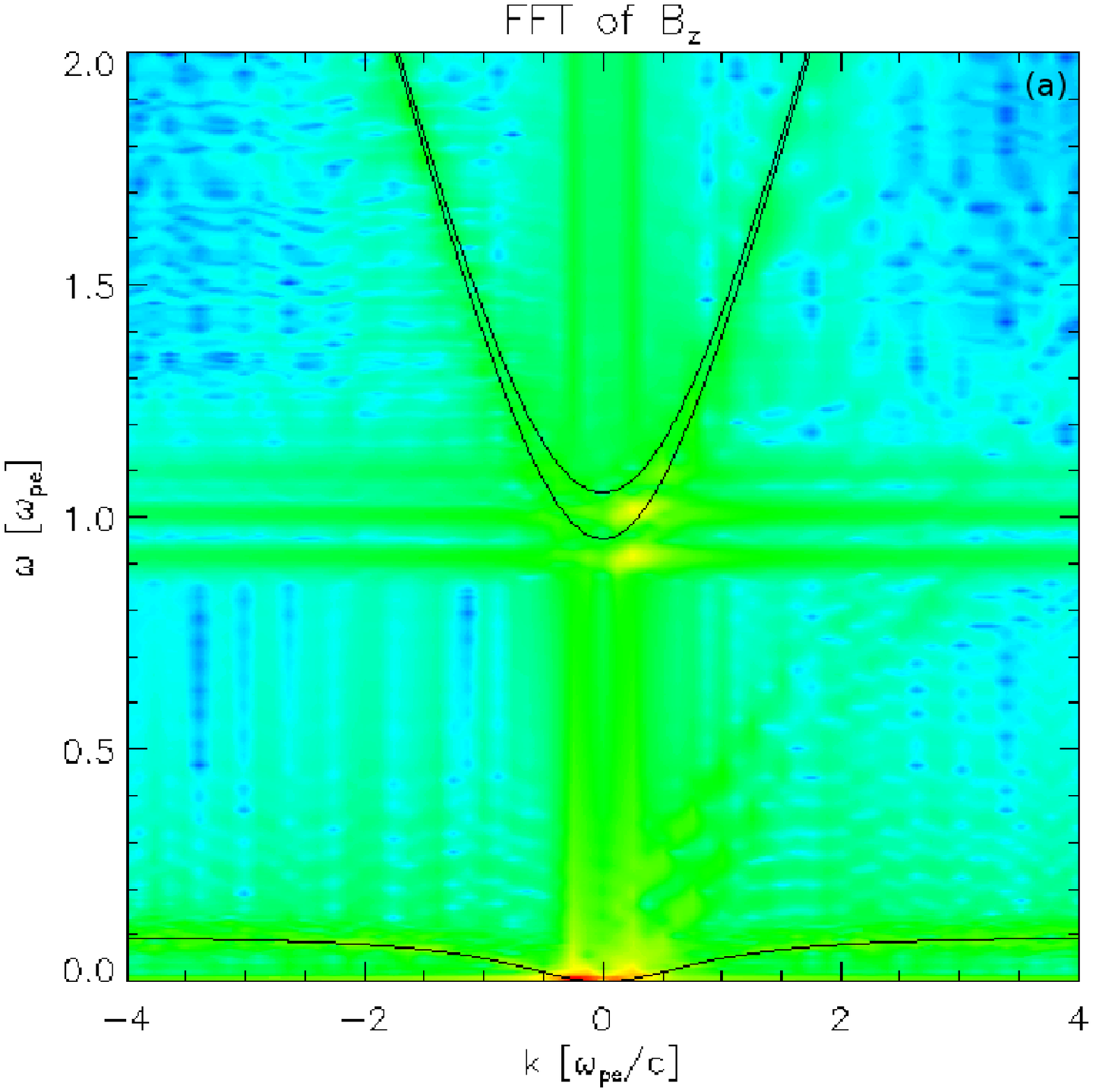} \\
\includegraphics[width=6.2cm]{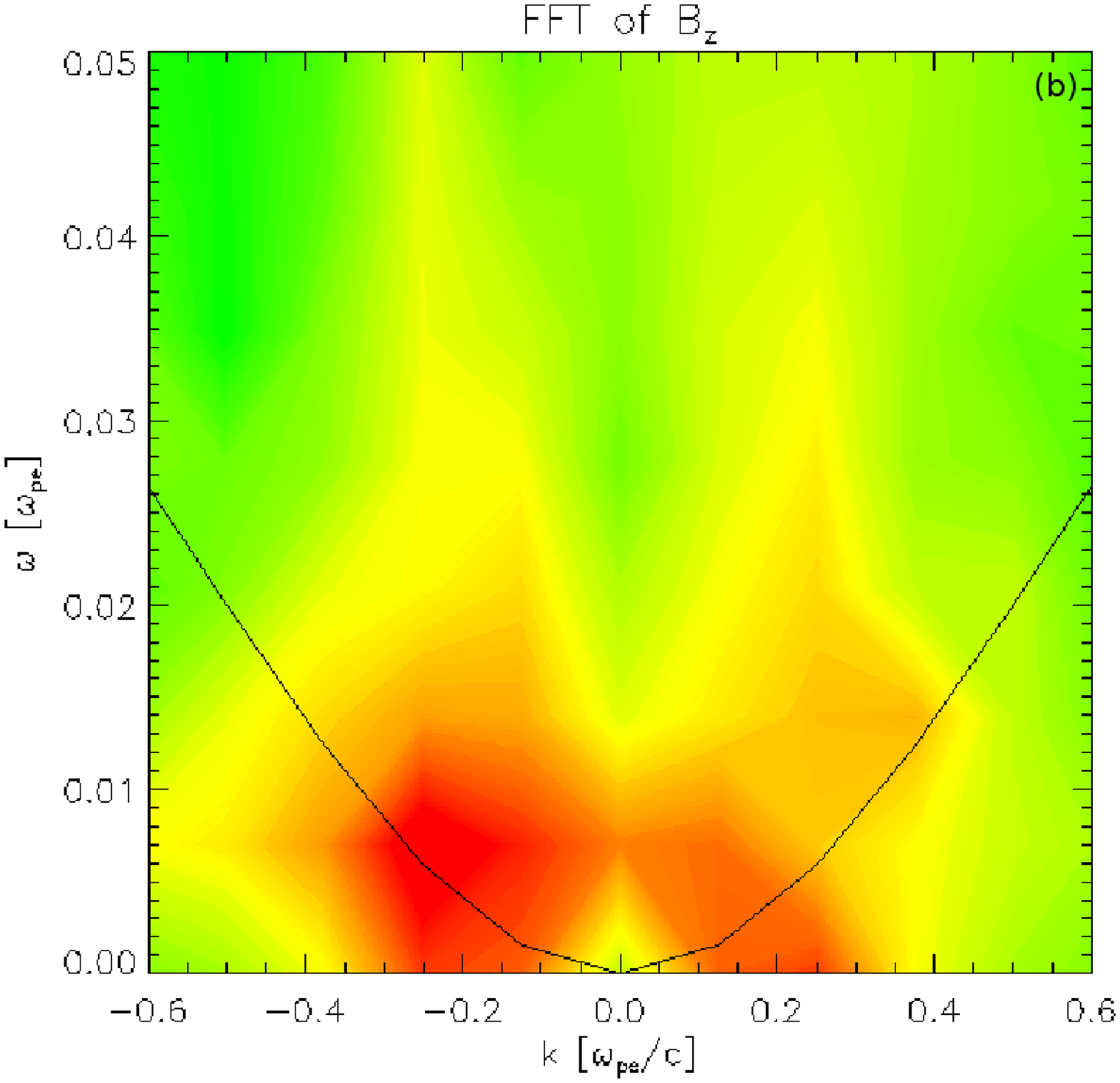} \\
\end{center}
\caption{
(a) FFT of the $B_z$ in the injection region of the reference run, with $\omega_{ce}/\omega_{pe}(x=0) = 0.1$,
depicted in the logarithmic colour scale and non equal steps. 
Two modes excited with slightly different frequency around $\omega_{pe}(x=0)$ are evident
in the upper part of the graph. 
The maxima lie at the intersection of the wavenumber prescribed by the length of the beam, 
$k_{beam} = 2 \pi / L_{beam}$, and the $0^{th}$ root of the dispersion curves of the L-- and R-- modes.
At $k_{beam}$ the frequency difference becomes somewhat smaller then $\omega_{ce}$.  
Dispersion curve of the $1^{st}$ root of the R$-$ mode is seen to have asymptotic at 
the $\omega_{ce}/\omega_{pe}(x=0)$, in accordance with the dispersion relation for the L--waves,
$k = \omega / c \cdot \{1 + \omega_{pe}^2 /(\omega ( \omega_{ce} - \omega  ) ) \}^{1/2}$ in Ref.\cite{Dendy}. 
(b) Enlarged lower part of the FFT graph from the top panel, depicting the area around whistler
curve in which the backwards travelling $\omega_{ce}$ harmonics exhibit strongly enhanced $B_{z}$. 
}
\label{fft_Bz_rr}
\end{figure}

\begin{figure}
\begin{center}
\includegraphics[width=6.2cm]{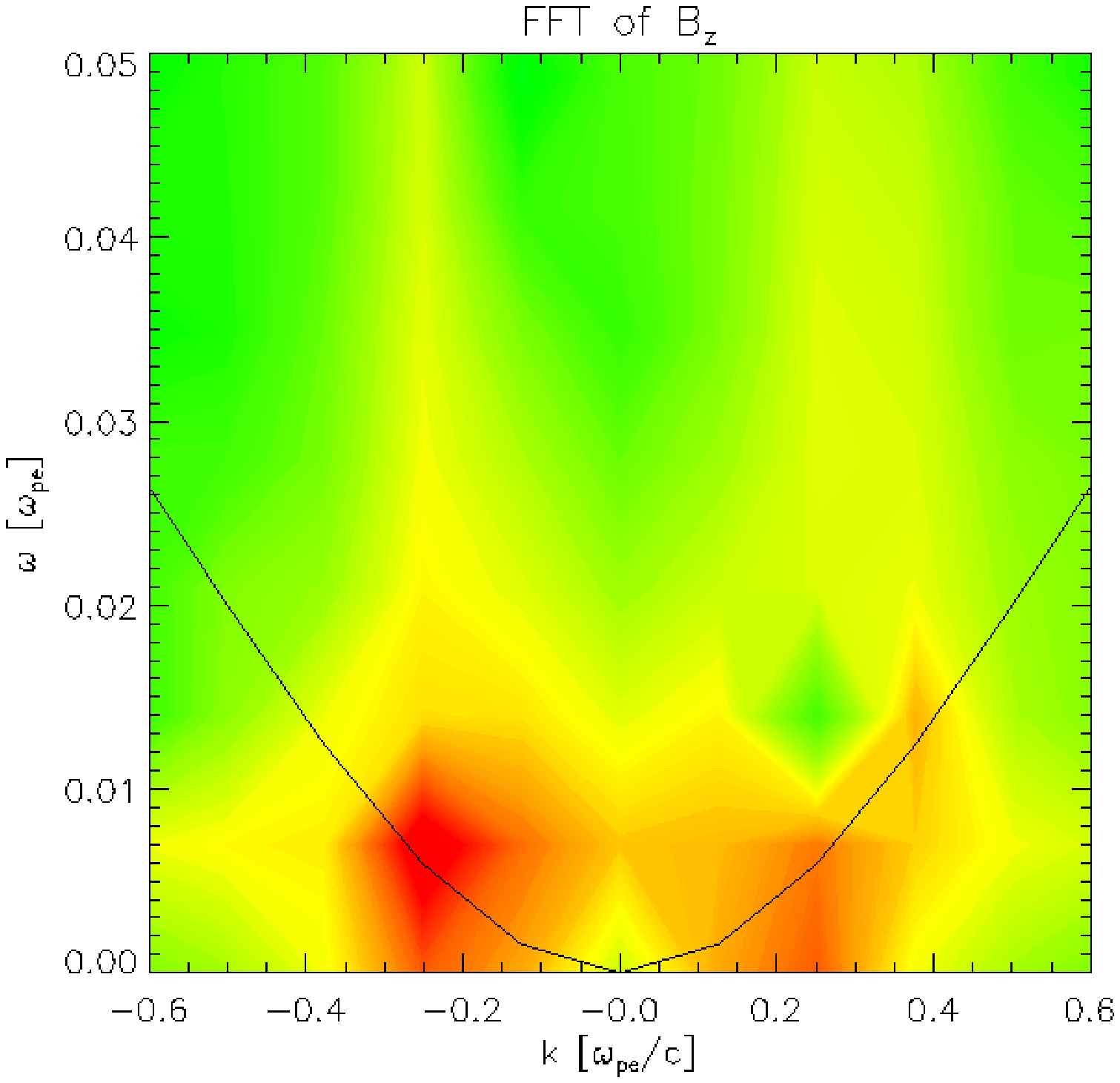} \\
\end{center}
\caption{ FFT of $B_{z}$ in the whistler wave enhanced area in the case when the two times narrower
beam than in the reference run was imposed. Overplotted black curve depicts the whistler wave dispersion 
relation. 
}
\label{fft_Bz_2xnarrower}
\end{figure}

\subsubsection{Characterisation of the wave}

Further insight into the wave activity of the system is gained through the Fast Fourier Transform (FFT)
analysis. Fig.~\ref{fft_Bz_rr} displays the two$-$dimensional FFT of the $B_{z}$ in the
injection region of the reference run, with the injection region being defined as the stripe
 $x \in [0, 50] $ and $t \in [0, t_{end}] $. Logarithmic colour scale is employed.
Additionally, in order to have better visual display, non equal steps of the colorscale are
applied. The smallest step is chosen to be five times smaller than the average
step and put at the largest value of the FFT($B_{z}$), because the function was seen to have significantly
more variations at its larger values.
The step increases uniformly for decreasing FFT($B_{z}$) and reaches its maximal value, five times
larger than the average step, on the side of the minimal value of FFT($B_{z}$). 
In a FFT graph the group velocity is inferred
as a tangent to the dispersion curves, $v_{g} = \partial \omega/\partial k$, 
and the phase velocity is a straight line passing through the origin, $v_{ph} = \omega/k$ .

Fig.~\ref{fft_Bz_rr}~(a) displays two initially excited high frequency wave modes. Overplotted black
parabolas are $0^{th}$ root of the dispersion relations for the electromagnetic R-wave (higher) and L-wave (lower) 
with electron plasma density at the left side of the simulation box, $\omega_{pe}(x=0)$, taken.
The black curve in the bottom part of the graph is $1^{st}$ root of the R-mode, known as the whistler wave
, see $e.g.$ Eqs.~(3.38) and (3.41) in Ref.~\cite{Dendy}.
Whistler wave dispersion curve approaches asymptotic value $\omega_{ce}/\omega_{pe}(x=0)=0.1$
at large $k$.
The maxima of the FFT($B_z$) in the plot lie at the intersection of $k_{beam} = 2 \pi /L_{beam}$
and the dispersion curves of the L and R modes. Therefore, the waves were identified as 
$L$ and $R$ waves by Schmitz and Tsiklauri \cite{2013PhPl...Schmitz}.
The frequency difference of the two modes at $k=0$, equal to $\omega_R(k=0) -\omega_L(k=0) = 
\omega_{ce}/2 \cdot \{ \sqrt(1+4\omega_{pe}^2/\omega_{ce}^2)+1 \}
- \omega_{ce}/2 \cdot \{ \sqrt(1+4\omega_{pe}^2/\omega_{ce}^2)-1 \}  = \omega_{ce}$ \cite{Dendy},
has a smaller  than $\omega_{ce}$ value at $k=k_{beam}$.
Superposition of the two modes with slightly different frequency
results in the beating of the field oscillations, which can be seen in the upper left insert of   
Fig.~\ref{Bz_rr}~(c). 
The formation of the beats and the backwards propagating wave studied here
takes place in the initial injection region of the beam, as seen in  
of Figs.~\ref{Bz_rr}~(a) and (b). The beats are formed by superposition of $\omega_L(k_{beam})$ and $\omega_R(k_{beam})$, 
$cf.$ Ref.~\cite{2013PhPl...Schmitz}. 
The formed wave packet starts moving from the injection region into 
the the lower density region. The
almost horizontal lines of the $E_{y}$ visible in the insert in the time-distance plot become inclined 
(Fig.~\ref{Bz_rr}~(c)) due to the wave refraction caused by the density gradient,
thereby indicating decrease of the phase velocity, $v_{ph}$, to finite values. 
The decrease of the phase velocity inferred in
Figs.~\ref{Bz_rr}~(a) and (c) (see also Fig.~5 from Ref.~\cite{2013PhPl...Schmitz} for its dynamic in time version) 
continues throughout  the region of
the decreasing density. At the time $\approx$ 500 $\omega_{pe}(x=0)$ it becomes almost equal 
to the speed of light, $v_{ph} = c$. 
The wave packet has eventually turned into a freely escaping electromagnetic wave,
as reported by Ref.~\cite{2013PhPl...Schmitz}, due to $(\omega,k)$$-$space drift, $i.e.$ wave refraction.

Fig.~\ref{fft_Bz_rr}~(b) presents enlarged lower part of the FFT graph from the top panel.
Whistler dispersion relation curve is overplotted with black curve. Enhanced magnetic field component $B_{z}$
evident along the whole whistler curve is seen to be strongly enhanced  at small $\omega$ and a small $k$.

Fig.~\ref{fft_Bz_2xnarrower} presents zoom into whistler region of the FFT of $B_y$ 
for the case when the beam is taken to have half of the width of the reference run. Comparison
with Fig.~\ref{Bz_rr}~(c)  leads to a  conclusion that the change of the width of the
beam does not play a role for whistlers.

\begin{figure}
\begin{center}
\includegraphics[width=6.2cm]{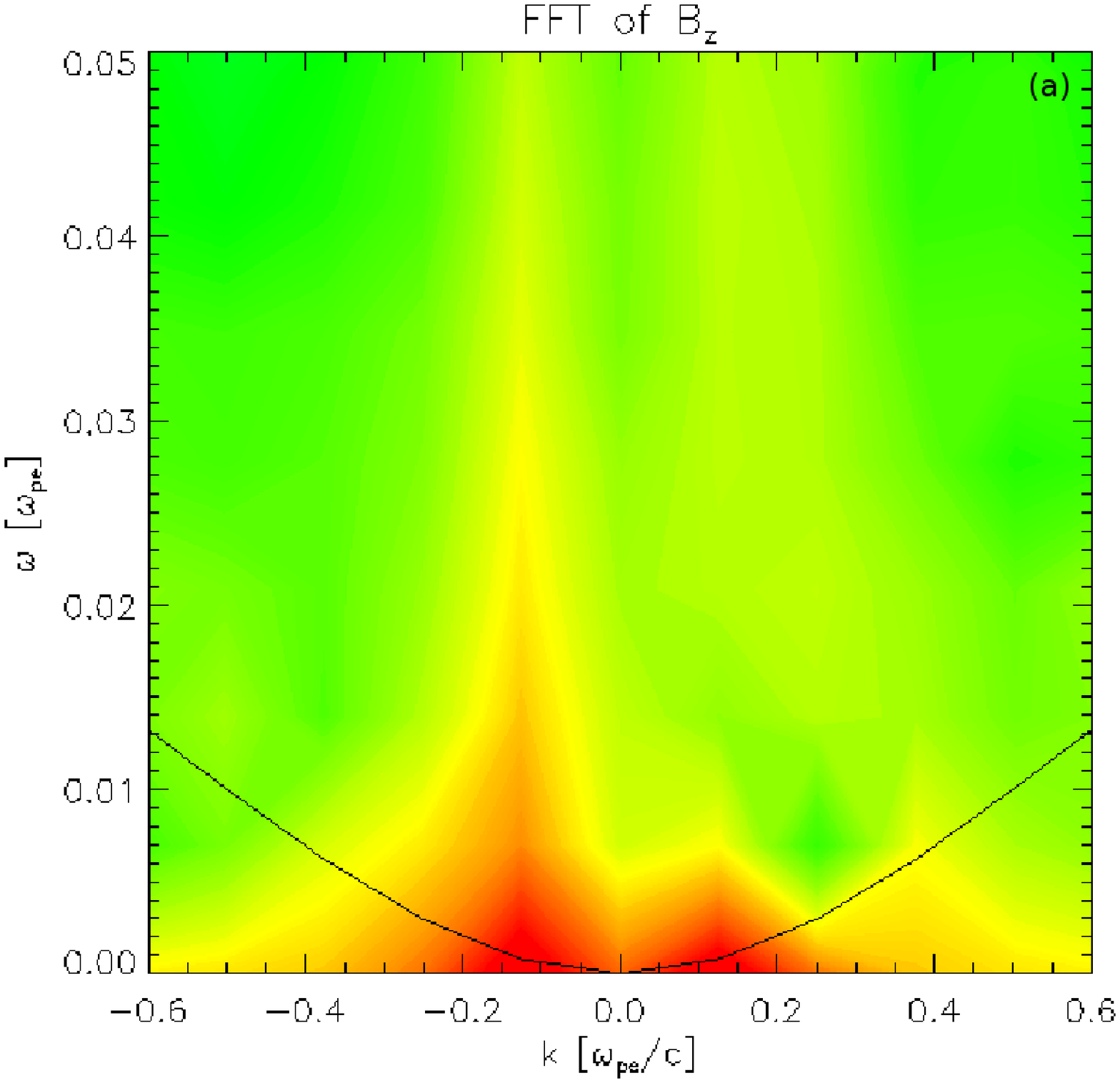} \\
\includegraphics[width=6.2cm]{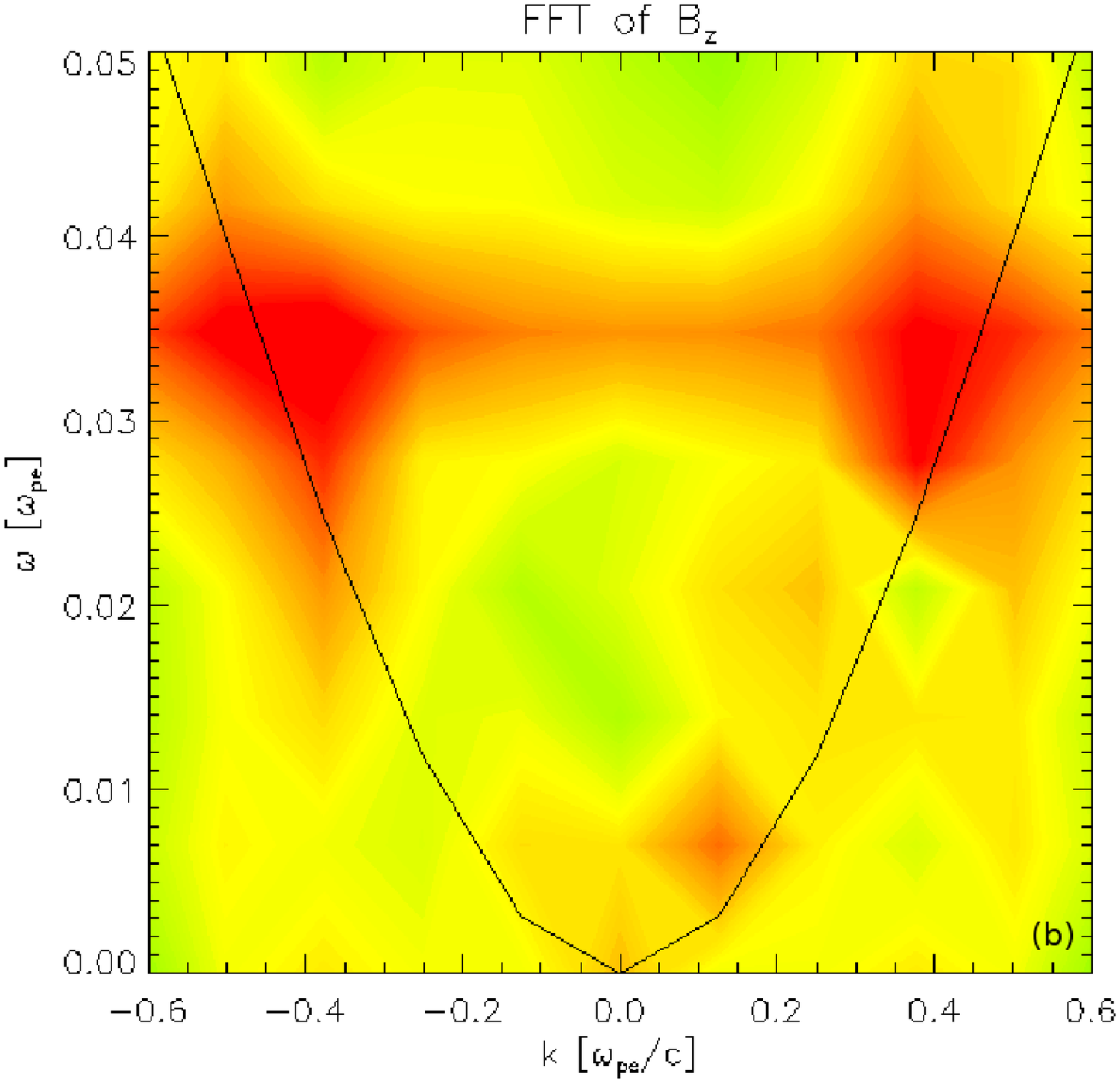} \\
\end{center}
\caption{(a) FFT graph for $B_z$ at prescribed $\omega_{ce}/\omega_{pe}($x=0$) = 0.05$.
Backwards travelling harmonics
have the strong enhancement of the $B_{z}$ at smaller angular frequencies $\omega$ and smaller wavenumbers $k$ 
than in the reference run.
(b) FFT graph for $B_z$ at prescribed $\omega_{ce}/\omega_{pe}($x=0$) = 0.2$.
Backwards travelling harmonics have enhancement of the $B_{z}$ at larger wavenumbers $k$ and
larger angular frequency $\omega$ than in the reference run.
}
\label{fft_Bz_modes}
\end{figure}

FFT graphs zoomed into the whistler area for the $B_{0x}$ taken to be
$2$ times smaller and $2$ times larger than in the reference run are presented 
at Figs.~\ref{fft_Bz_modes}~(a) and (b), respectively. 
In all the examined cases (applied $\omega_{ce}$, 1/2$\cdot$$\omega_{ce}$
and 2$\cdot$$\omega_{ce}$)) the initially excited backward propagating harmonics  
exhibit enhancements of the $B_{z}$  along the whistler dispersion curves  
and are therefore conclusively identified as whistlers.

\subsection{Excitation of the whistler waves}

\begin{figure}
\begin{center}
\includegraphics[width=7cm]{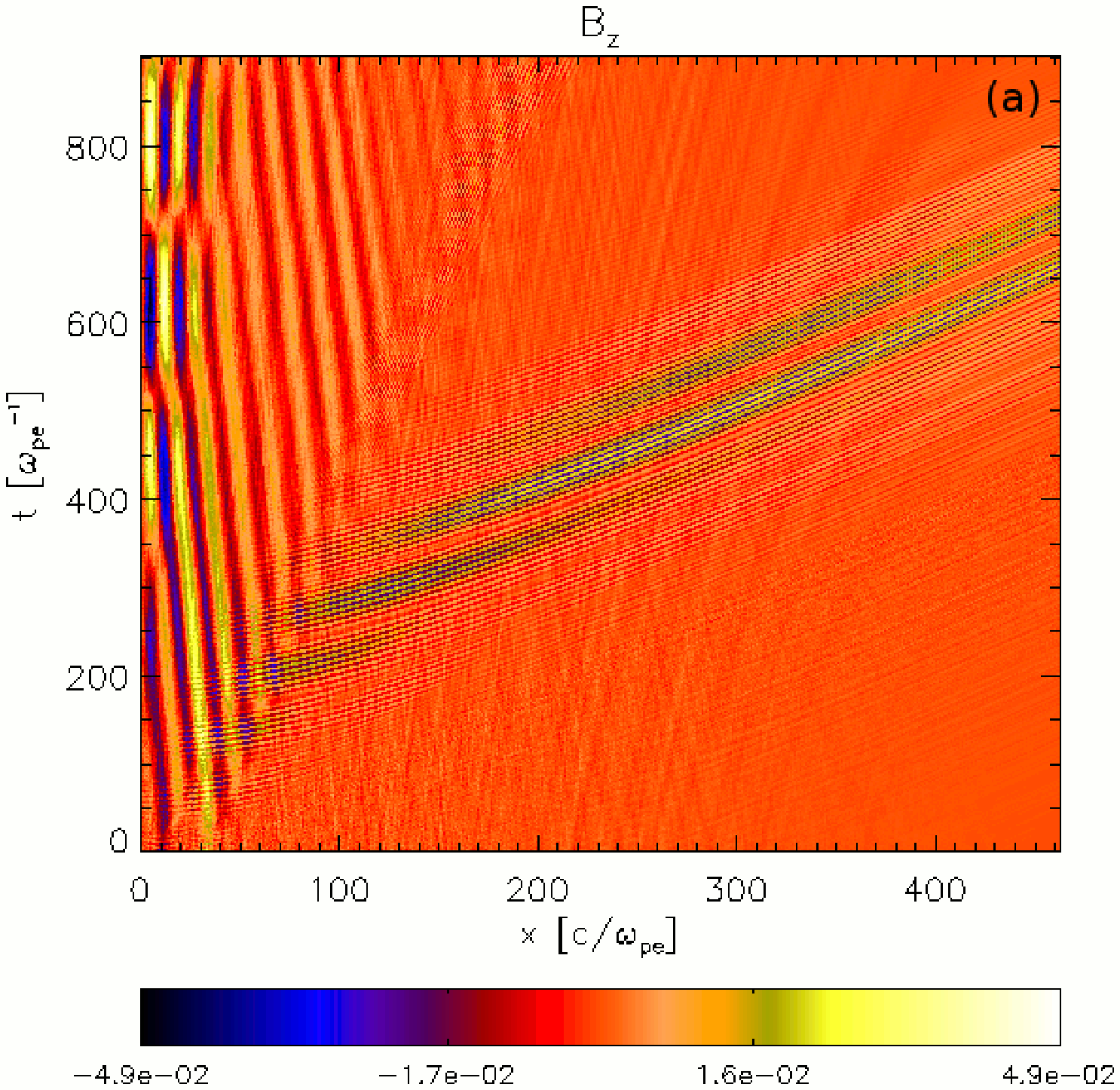} \\
\includegraphics[width=6.2cm]{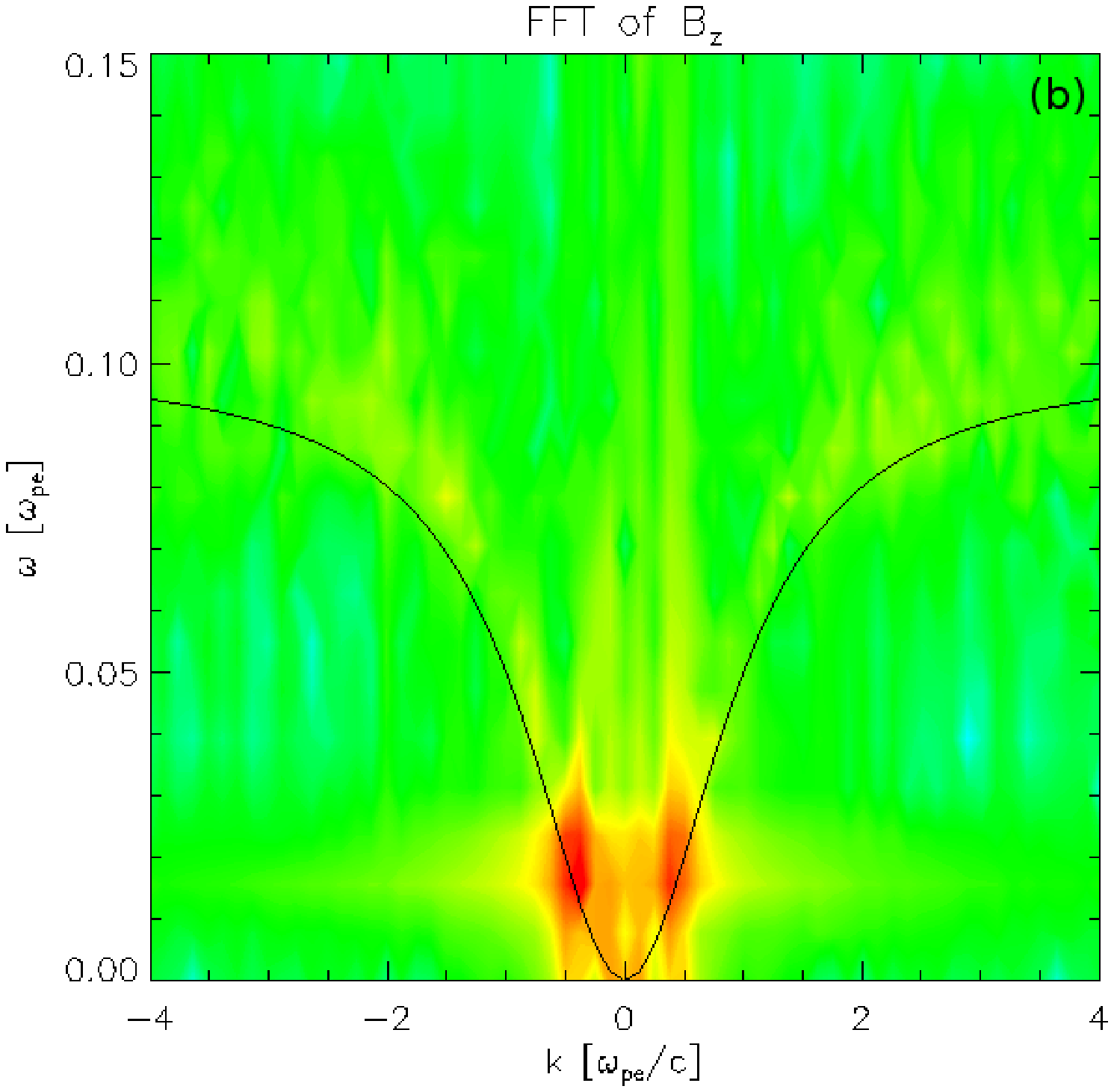} \\
\includegraphics[width=7cm]{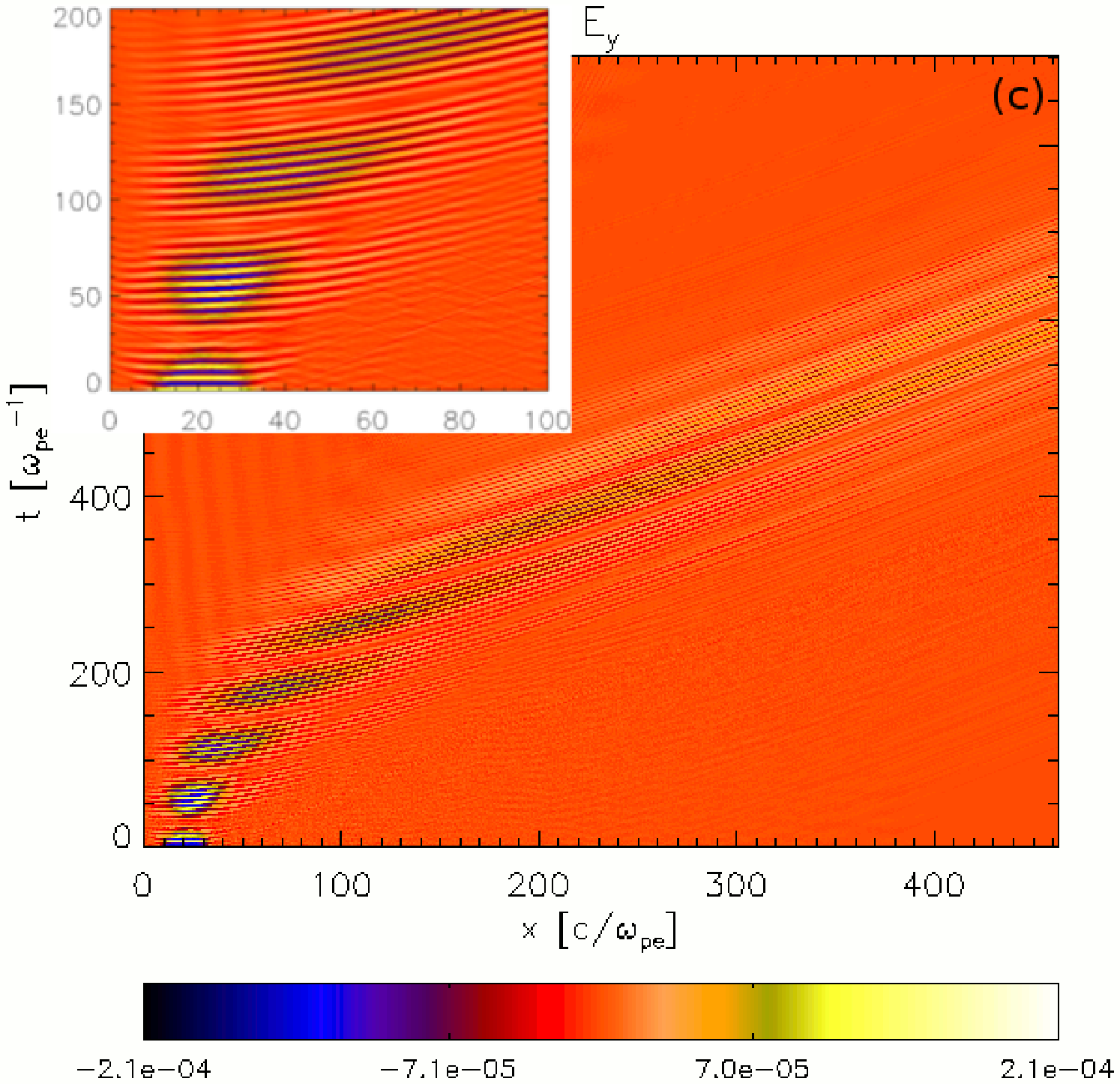} \\
\end{center}
\caption{(a) Time--distance plot of $B_z$ for the case when the beam is two times slower than
in the reference run. The backwards propagating harmonics are evident in the
wake--side of the beam.
(b) FFT of the $B_z$ enlarged in the whistler area.
The enhancement of the $B_z$ along the whistler dispersion curve is evident.
The strong enhancement of $B_z$ takes place at larger $k$ and $\omega$ than in the reference run.
(c) Time--distance plot of $E_y$ for the same run.
}
\label{fft_Bz_slow2}
\end{figure}

\begin{figure}
\begin{center}
\includegraphics[width=7cm]{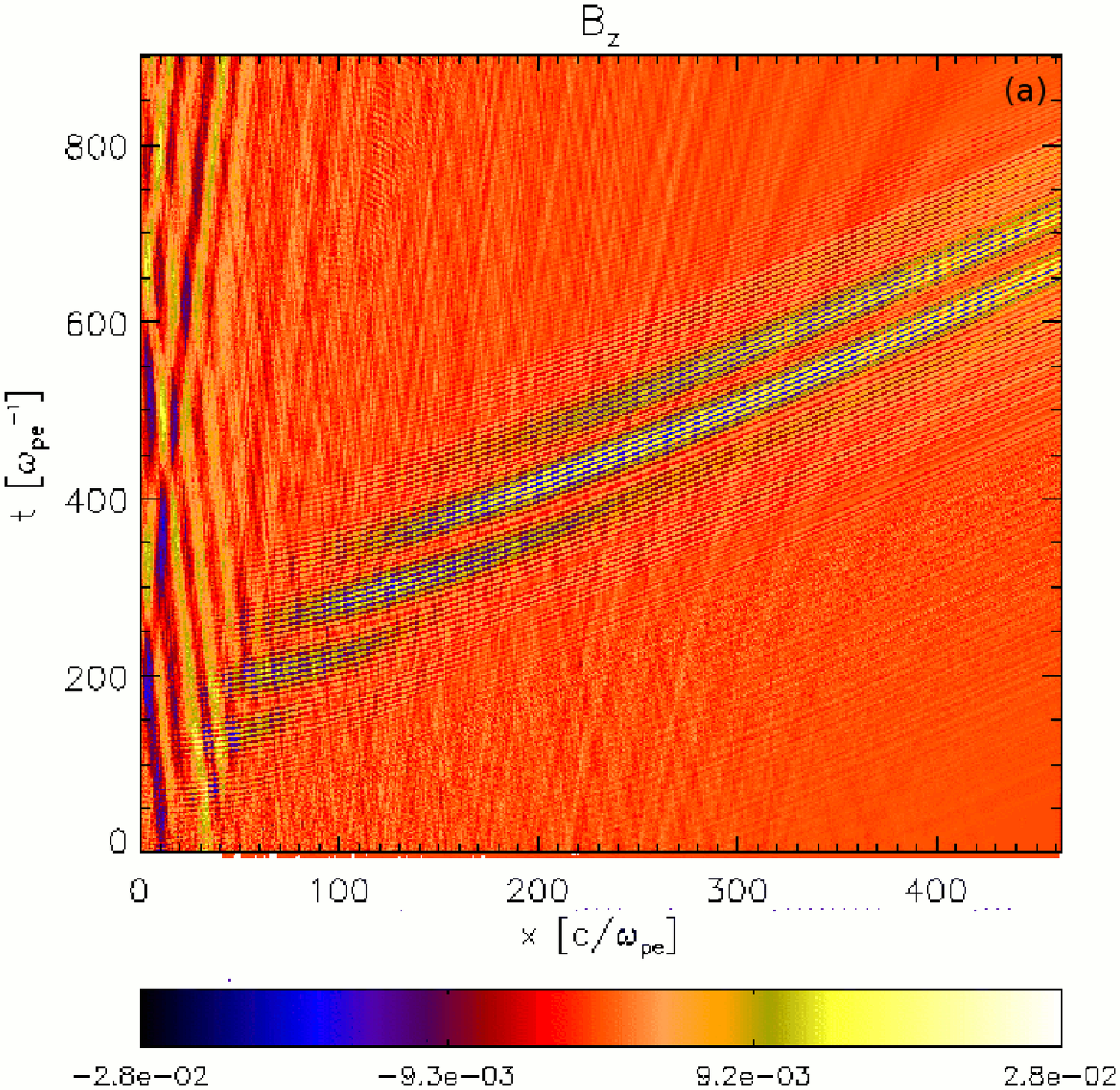} \\
\includegraphics[width=6.2cm]{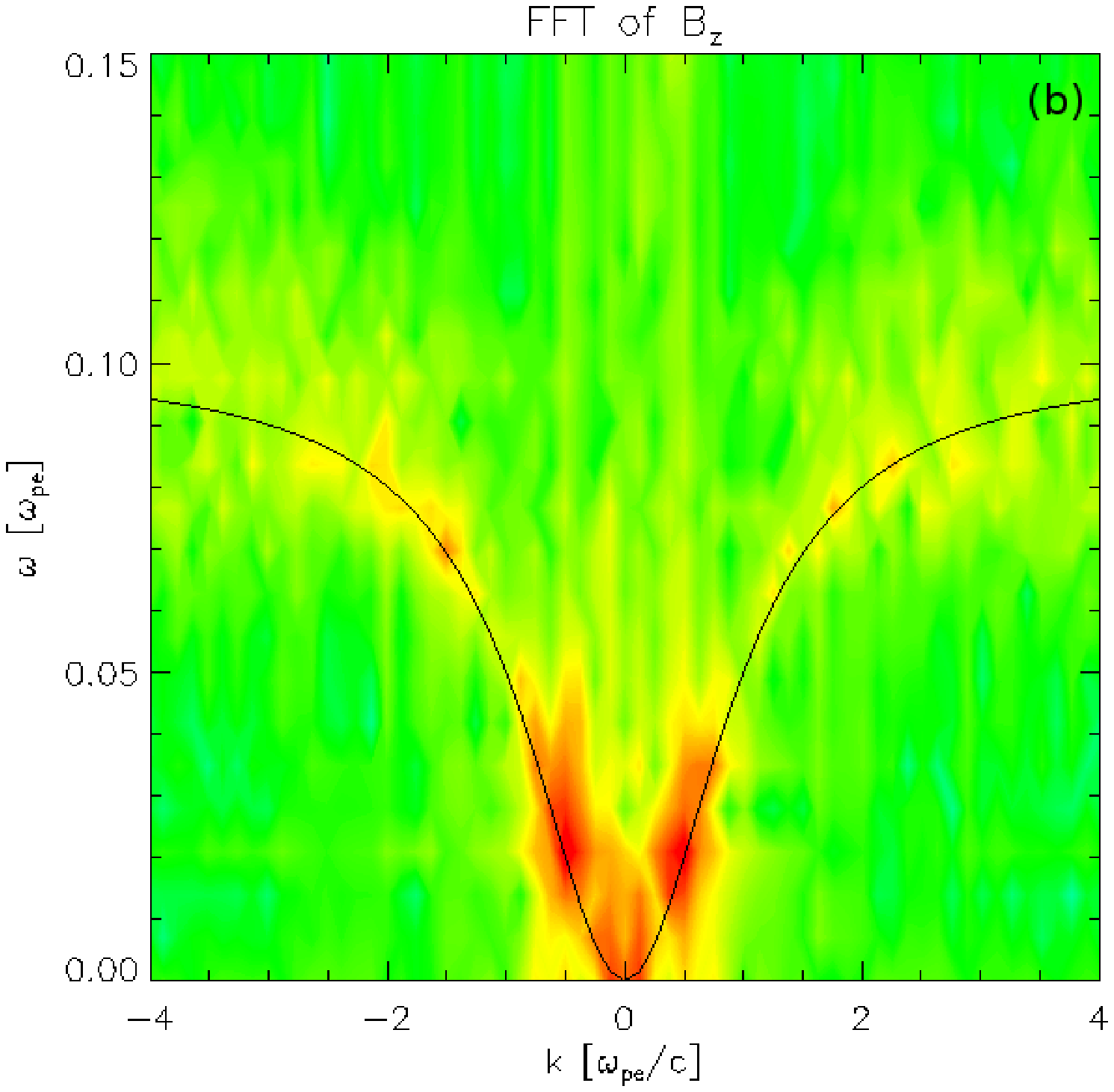} \\
\includegraphics[width=7cm]{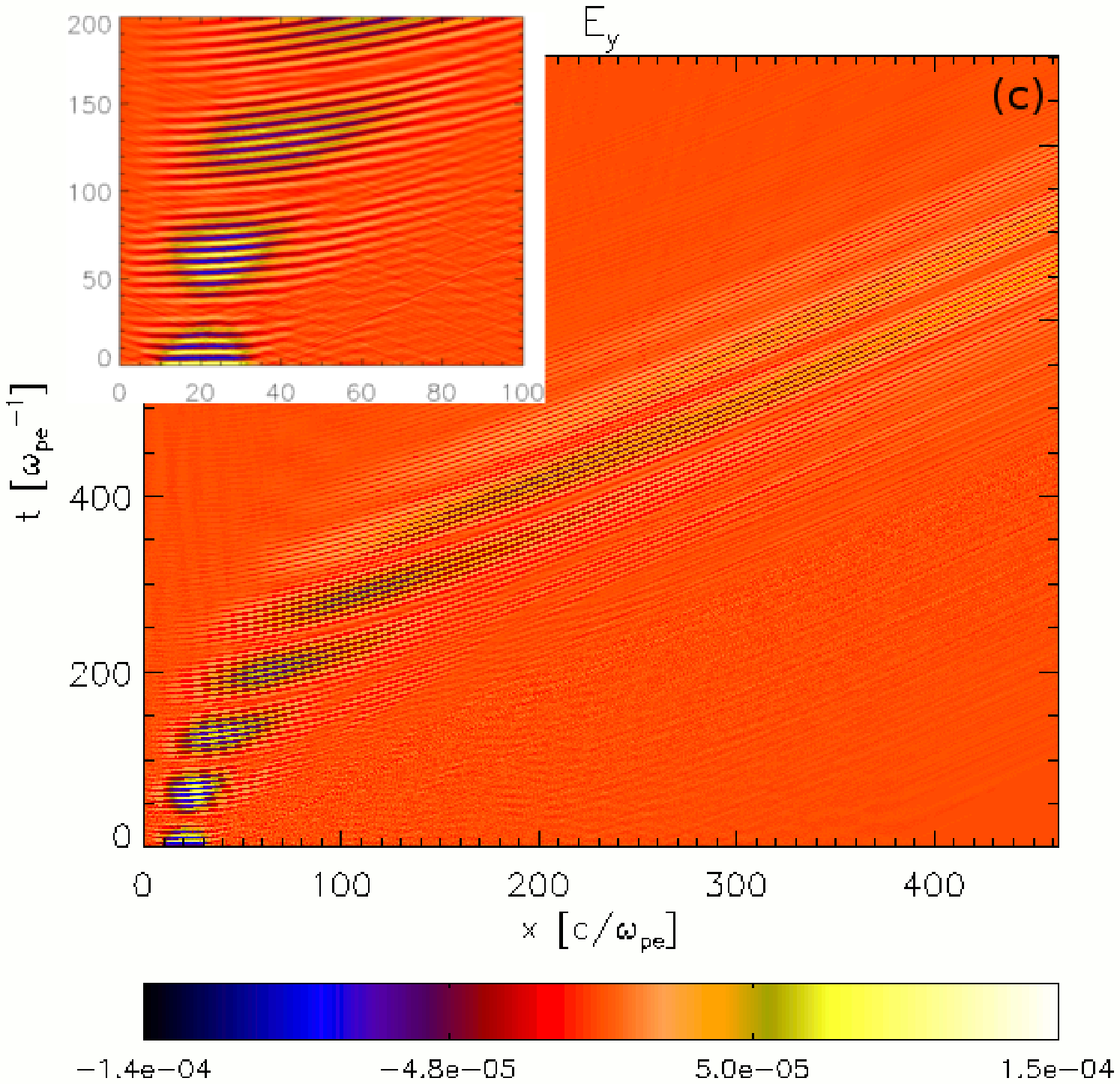} \\
\end{center}
\caption{(a) Time--distance plot of $B_z$ for the case when the beam is three times slower than
in the reference run. A clear pattern of backwards propagating harmonics is evident only
in the injection region.
(b) FFT of the $B_z$ enlarged in the whistler area.
The enhancement of the $B_z$ is more spread to larger wavenumbers. The strongest enhancement takes
place also at $k$ and $\omega$ bigger than in the reference run.
(c) Time--distance plot of $E_y$ for the same run.
}
\label{fft_Bz_slow3}
\end{figure}

In order to determine how the backwards propagating whistlers are generated, 
the following dispersion relation  for whistlers is employed \cite{SteZaiNak12}:
\begin{equation} \label{whistler_dispersion}
\omega = \omega_{ce} \frac{k^2 c^2 \mid cos(\vartheta) \mid}{\omega_{pe}^2 + k^2 c^2} \\,
\end{equation}
where $\theta$ stands for the angle between $\vec k$ and $\vec B$. Further, relativistic
Doppler-shifted cyclotron resonance condition is applied:
\begin{equation} \label{cyclotron_resonance}
\frac{s}{\gamma} \omega_{ce} = \omega - \vec k_{\parallel} \cdot  \vec v_{beam,\parallel}  \\,
\end{equation}
where $s$ represents particular cyclotron harmonic and
$\gamma$ is the Lorentz factor, $\gamma = (1 - v_{beam,\parallel}^2/c^2)^{-1/2} \approx 1.069 $. 
The resonant beam electrons drift in the opposite direction as 
the whistlers, thereby yielding the term 
$\vec k_{\parallel} \cdot \vec v_{beam,\parallel}$ to be negative.  
Whistler mode waves exist in the range $0 < \omega < \omega_{ce}$. Therefore, right hand side
of Eq.(\ref{cyclotron_resonance}) is positive and, consequently, the wave harmonic number $s$ is
positive, $s > 0$. 
Relativistic correction accounted for in Eq.(\ref{cyclotron_resonance}) is
found to play a decisive role in establishing the point of resonance between the electrons and the whistler waves,
analogously to Ref.~\cite{1979ApJ...Wu_Lee}.
Combination of Eqs. (\ref{whistler_dispersion}) and (\ref{cyclotron_resonance})
yields the equation for the velocity of the  relativistic beam electrons in the Doppler-shifted resonance
 with the whistler waves:
 \begin{equation} \label{electron_velocity}
\mid v_{beam,\parallel} \mid  = \frac{c}{\omega_{pe}} \mid \frac{s}{\gamma}\omega_{ce} - \omega \mid
 \sqrt{\frac{\omega_{ce}\mid cos(\vartheta) \mid - \omega}{\omega} } \\,
\end{equation}
which is rewritten as a cubic equation in $\omega$:
\begin{multline} \label{eq:cubic}
\omega^{3} - \omega^{2} \cdot \omega_{ce} (\mid cos(\vartheta)  \mid + 2\frac{s}{\gamma}) \\ 
+ \omega [ v^{2}_{\parallel} \frac{\omega_{pe}^2}{c^2} 
+ \frac{s}{\gamma} \omega_{ce}^{2} (2\mid cos(\vartheta)  \mid + \frac{s}{\gamma} ) ] \\
- \frac{s^2}{\gamma^2} \omega_{ce}^3 \mid cos(\vartheta) \mid = 0 . 
\end{multline}
The imposed initial condition specifies the ratio $\omega_{ce}/\omega_{pe}($x=0$)$, where
whistler dispersion curve has the asymptotic,
$\omega/\omega_{pe}($x=0$) = \omega_{ce}/\omega_{ce}($x=0$)$, 
as seen $e.g.$ in Fig.~\ref{fft_Bz_rr}~(a). Higher order cyclotron resonances,
$s>1$, are found to yield solutions above the asymptotic. The only acceptable
solution is found to be at the $ s = 1 $, which is normal Doppler-shifted relativistic resonance. 
Resonance patterns apparent in the FFT graphs, closely related to the harmonics $s=1$,
are presented below.

In the reference run  the prescribed ratio is $\omega_{ce} / \omega_{pe}(x=0) = 0.1$
The root of Eq.(\ref{eq:cubic}) yields $\omega_{0}(s=1)/\omega_{pe}(x=0) = 0.0058$. 
This finding is closely resembling the resonance apparent in the FFT graph of the  $B_{z}$ component 
at $k \approx 0.25$ presented in Fig.~\ref{fft_Bz_rr}~(b), that clearly sits on 
the whistler dispersion curve.

When $\omega_{ce} / \omega_{pe}(x=0) = 0.05$ 
the solution of Eq.(\ref{eq:cubic}) gives $\omega_{0}(s=1)/\omega_{pe}(x=0) = 0.00083$. 
This corresponds well with resonance at the $k \approx 0.09$ seen on  
Fig.~\ref{fft_Bz_modes}~(a). 
In this particular case the smallest normalised angular frequency that can be resolved is
$\omega_{min}/\omega_{pe}(x=0)=0.0011$, which is larger than the value we calculated.
The choice of the stripe boundaries in the time-distance plot determines the minimal normalised
wavenumber to be $k_{min}=0.1257$, which is larger from the wavenumber determined from the 
whistler dispersion relation. Therefore, in this case it is not possible to see 
the exact value of the calculated resonance value in the FFT graph.

In the case $\omega_{ce} / \omega_{pe}(x=0) = 0.2$ the solution of Eq.(\ref{eq:cubic}) yields
$\omega_{0}(s=1)/\omega_{pe}(x=0) = 0.0314$. The solution
corresponds reasonably well with resonant enhancements in the FFT graph apparent 
at $ k \approx 0.42 $, as depicted in Fig.~\ref{fft_Bz_modes}~(b),
which, again, closely follow whistler dispersion relation.

In order to support the normal relativistic Doppler-shifted relativistic resonance 
as the mechanism which excites whistlers in the wake--side of the beam, 
velocity of the beam is altered to $c/4$ and $c/6$, as presented in 
Figs.~\ref{fft_Bz_slow2} and \ref{fft_Bz_slow3}, respectively.
Figs.~\ref{fft_Bz_slow2}~(a) and \ref{fft_Bz_slow3}~(a)  display the time$-$distance graph for 
the magnetic field  component $B_z$ for the $v_{beam}$ equal to $c/4$ and $c/6$, respectively. 
On the wake side of the beam propagating with $v_{beam} = c/4$ the backwards propagating waves
are apparent. In the case of the $v_{beam} = c/6$ they are pronounced only in the injection region. 

Figs.~\ref{fft_Bz_slow2}~(b) and \ref{fft_Bz_slow3}~(b) display the FFT of 
the magnetic field  component $B_z$ for
the case of the $v_{beam}$ equal to $c/4$ and $c/2$, respectively.
The enhanced $B_z$ is seen to be more pronounced along the whistler curve also at the larger 
wavenumbers $k$. 
For the case of the beam velocity $v_{beam} = c/4$ the solution of Eq.(\ref{eq:cubic}) is 
$\omega_{0}(s=1)/\omega_{pe}(x=0) = 0.0173$. 
This is reasonably well confirmed in Fig.~\ref{fft_Bz_slow2}~(b) 
at $k \approx 0.44$ in the resonant area.
In the case of the $v_{beam} = c/6$ the $\omega_{0}(s=1)/\omega_{pe}(x=0) = 0.0272$.
This corresponds well with the resonant areas in the FFT of $B_{z}$ graph  presented 
in Fig.~\ref{fft_Bz_slow3}~(b) at $k \approx 0.57$. In the case of the slower beam
there is also more intensity of the magnetic field component $B_{z}$ spread along the whistler
curve, $i.e.$ also at larger frequencies and wavenumbers.

The width of the beam does not enter Eq.(\ref{eq:cubic}) and, as evidenced
in Fig.~\ref{fft_Bz_2xnarrower}, does not influence the excitation of the whistlers.
On the other hand, change of the beam velocity
enters Eq.(\ref{eq:cubic}) as $v_{beam,\parallel} = v_{beam}/\sqrt2$ and
the yielded solutions, which are the Doppler-shifted relativistic resonance solutions,
are found to correspond well to the inferred strong field enhancements.

The backward propagating whistler mode waves are found to be excited in the injection region
already at the initial time.
The whole electron beam is placed in the simulation box at the
beginning. As the beam does not cross any boundary, no transition radiation is observed, as in 
Ref.~\cite{1999PhPl....6.1427S}. 
The beam has approximately three orders of magnitude smaller particle number density than the 
ambient plasma in the injection region, $x\in[0,50]$ $[c/\omega_{pe}(x=0)]$,
as seen in Fig.~\ref{beam_rr}. Such a beam is 
not dense enough to form a virtual cathode, as in Ref.~\cite{1987JGR....92.7673P}, 
which would result in the beam modulation such as observed in Ref.~\cite{2000ITPS...28..367S}.  
The beam in the simulation has very small density fluctuations at the initial time,
which grow during the simulation and, due to quasilinear relaxation,
 develop into the spiky density profile at the end, as seen in Fig.~\ref{beam_rr}.  
Slight beam particle density variation in the beam particle temporal profile, less than $10\%$, 
is apparent only when the beam reaches
$x=57$ $[c/\omega_{pe}(x=0)]$ and from that time on the variation of the beam particle density is seen 
to increase. However, whistlers are evident already in the injection region, 
$x\in[0,50]$ $[c/\omega_{pe}(x=0)]$,
where there is no sign of the beam density variation. 
Compressibility of the whistler mode can yield the whistlers additionally perturbing the beam density.
Thus we conclude that the whistler mode waves are occuring due to the
normal Doppler-shifted resonance with the drifting relativistic beam electrons.

\subsection{Partition of the energy}

\begin{figure}
\begin{center}
\includegraphics[width=6.2cm]{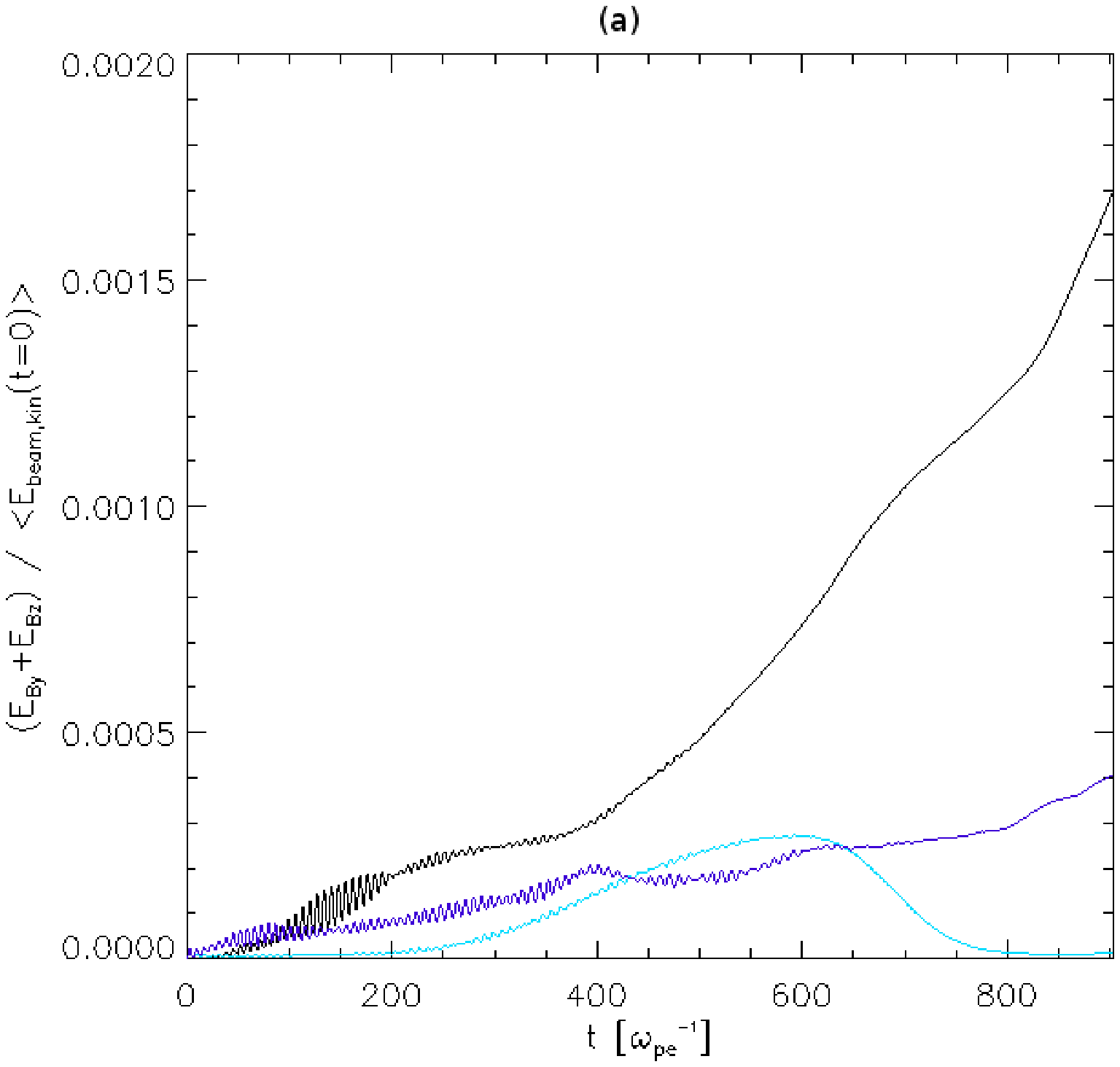} \\
\includegraphics[width=6.2cm]{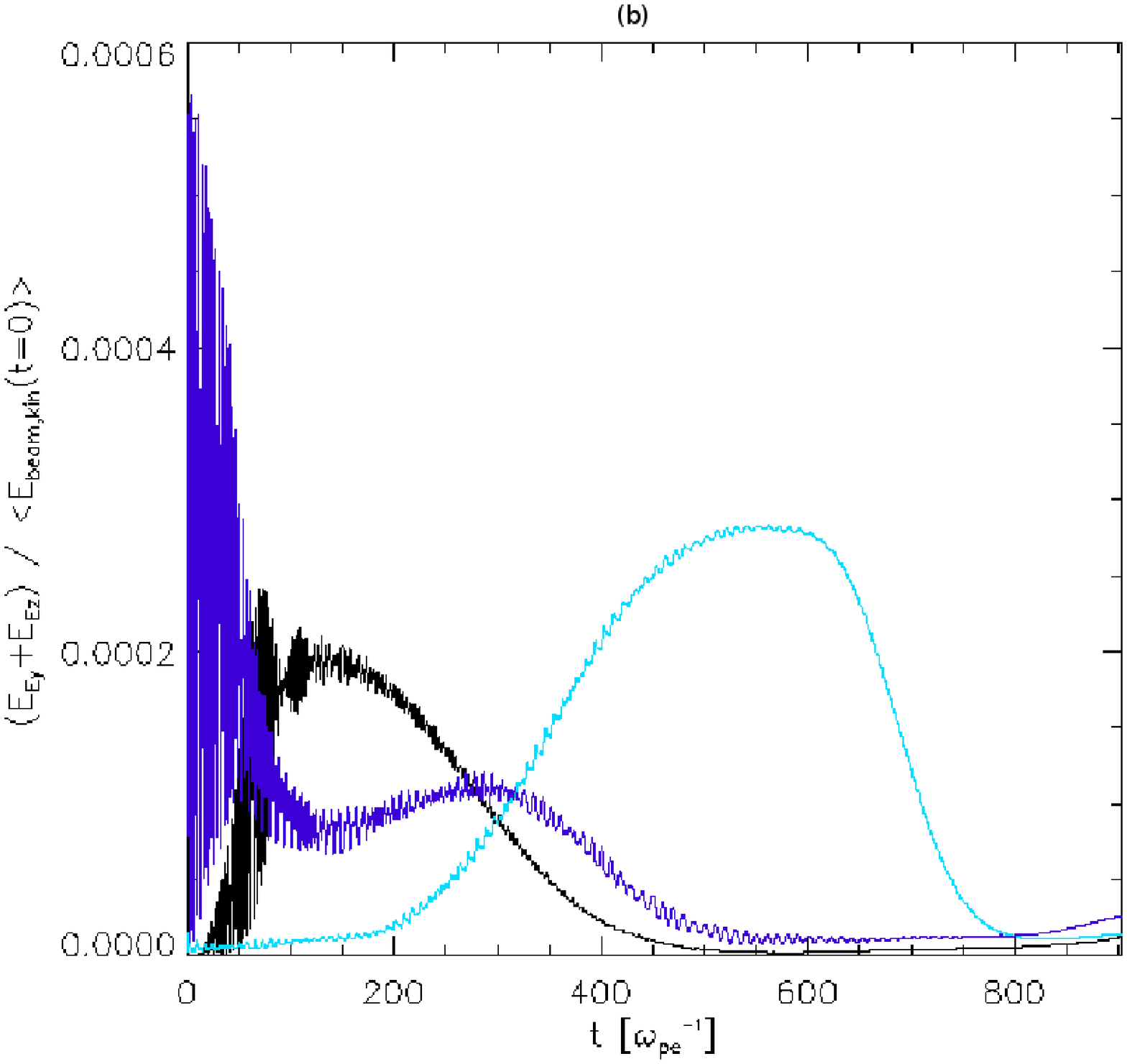} \\
\includegraphics[width=6.2cm]{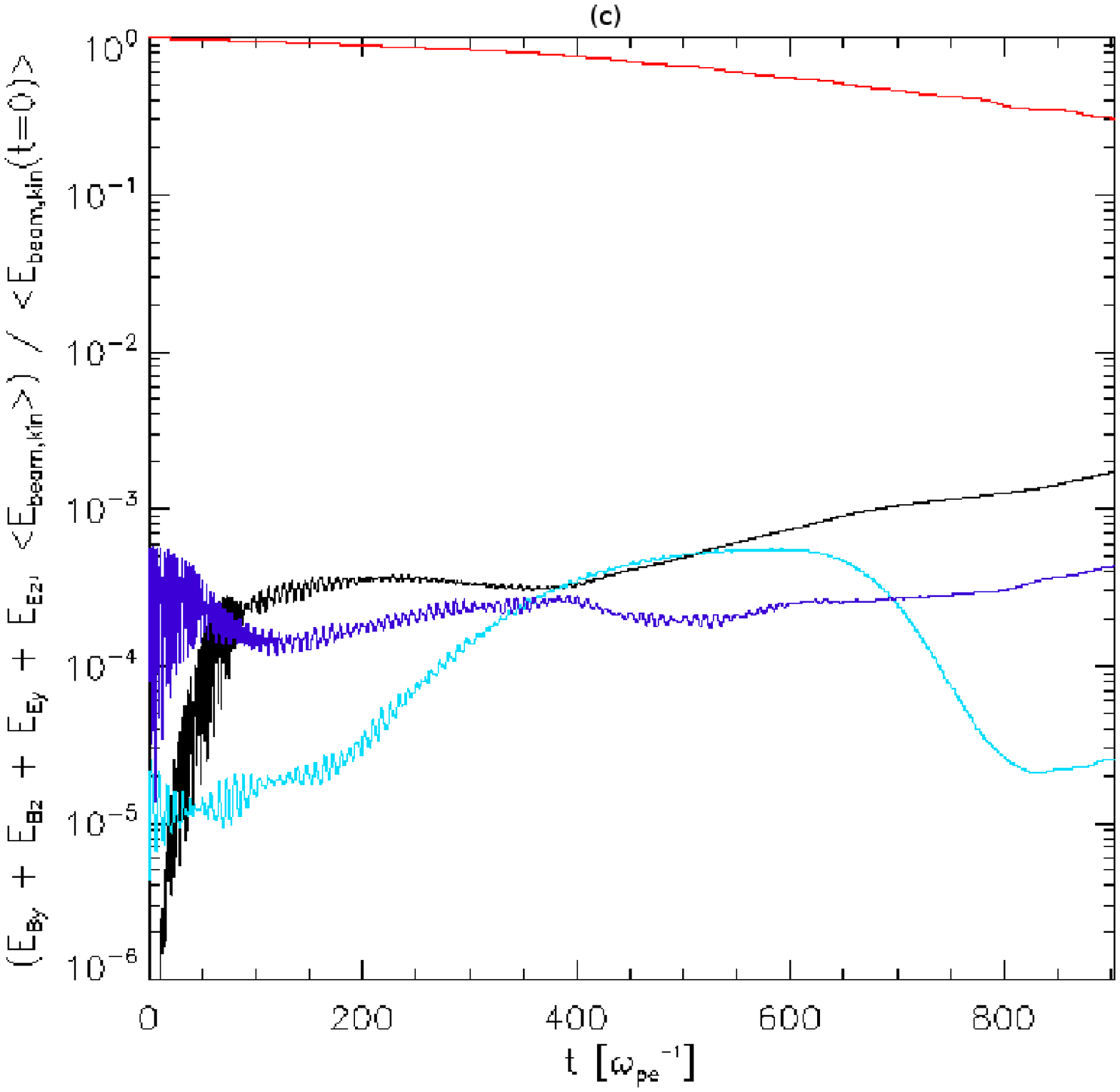} \\
\end{center}
\caption{
(a) Energy partition in the perpendicular components of the magnetic field, $B_{y}$ and $B_{z}$, 
normalised with the initial kinetic energy of the beam as a function of time. 
(b) Energy partition in the perpendicular components of the electric field, $E_{y}$ and $E_{z}$, 
normalised with the initial kinetic energy of the beam as a function of time.
(c) Partitioned total electromagnetic energy of the perpendicular components, 
$B_{y}$, $B_{z}$, $E_{y}$, $E_{z}$, and the beam kinetic energy (uppermost, decreasing, curve) 
normalised with the initial kinetic energy of the beam as a function of time.
At its peak the perpendicular elecromegnetic energy of the generated electromagnetic radiation is seen to be
$\approx 0.054 \%$ of the initial beam kinetic energy.
}
\label{energije}
\end{figure}

Analysis of the electromagnetic energy of the system is performed starting from the 
Poynting's theorem for electromagnetic energy as a function of time:
\begin{equation} \label{eq:Poyntings_theorem}
w(t) = \int \Big[\frac{1}{2}\varepsilon_{0}\vec E(x,t)^2 +
 \frac{1}{2\mu_{0}}\vec B(x,t)^2\Big]\mathrm{d}x \\,
\end{equation}
where $\varepsilon_{0}$ stands for the electric permittivity of the vacuum, $\mu_{0}$ for the magnetic
permeability of the vacuum and $\vec E$ and $\vec B$ are the electric and the magnetic field, respectively.
The electric and magnetic field can be decomposed into the Cartesian components:
\begin{equation} \label{eq:Cartesian_components}
 \begin{matrix}
  \vec E(x,t) = \vec E_{x}(x,t) + \vec E_{y}(x,t) + \vec E_{z}(x,t)  \\
  \vec B(x,t) = \vec B_{x}(x,t) + \vec B_{y}(x,t) + \vec B_{z}(x,t)  
 \end{matrix}
\end{equation}
The $x$ component of the electric field is contributing to the electrostatic energy, while the 
$x$ component of the magnetic field after subtraction of the prescribed $B_{0x}$ is in the order of 
numerical noise in our numerical simulations.
The perpendicular components, $y$ and $z$, of the electric and magnetic fields are contributing both 
to the whistler and electromagnetic wave energies. 

At every given instant of time the position of the beam is calculated and the time--distance plot
 is divided into three parts:
A part occupied by the whistlers, from left boundary of the numerical box to the back edge of the beam, a part
occupied by the beam, stretching between the front and the back edges of the beam, and a part in which
the freely escaping electromagnetic radiation is found, from front edge of the beam till the right boundary
of the box. 
The initial beam kinetic energy is calculated following Ref.~\cite{2012PhPl...Pechhacker} and found to be
$\left<E_{beam, kin}(t=0)\right> = 9.94607 \cdot 10^{-3}J$. Calculated perpendicular electromagnetic energy
components for the reference run 
are divided by the initial kinetic energy of the beam and presented in Fig.~\ref{energije}.
Whistlers are seen to originate at the front edge of the beam, however the beats (created by the superposition of
L-- and R-- electromagnetic waves) leading to the freely
escaping electromagnetic radiation are taking place in the region of the beam or even slightly behind it.
Thereby, the perpendicular electromagnetic energy contained in the beam is presented separately. 
Black curve depicts the energy in whistlers, cyan is for the energy of the electromagnetic radiation and 
blue curve depicts the fraction of the perpendicular magnetic energy in the beam (color online).
Fig.~\ref{energije}~(a) presents the perpendicular magnetic field energy as a function of time.
The fraction of the energy taken by the whistlers is seen to be largest and increasing with time.
The part of the perpendicular magnetic energy contained in the beam is smaller and slightly increasing with the time.
Electromagnetic waves are found to contain very little of the perpendicular magnetic energy initially.
Their energy is seen to grow up to a limit, and then to decrease due to the distribution becoming a ring as established
in Fig.~4 of Ref.~\cite{2013PhPl...Schmitz}.
Fig.~\ref{energije}~(b) presents the partitioned perpendicular electric field energies as a 
function of time. In the perpendicular electric field energy the whistlers are seen to take less of the
energy than the electromagnetic part. However, the whistlers contain substantially more energy in the
perpendicular magnetic than in the perpendicular electric field, approximately $70$ times more by the end of the simulation.
Therefore, the total perpendicular electromagnetic energy budget, as presented in Fig.~\ref{energije}~(c),
shows that the largest fraction of the energy is taken by the whistler waves.
The whistlers, taking $\approx 58 \%$ of the perpendicular electromagnetic energy, 
are consuming only $0.096 \%$ of the initial beam kinetic energy. 
Decrease of the beam kinetic energy divided by the initial beam kinetic energy in time is presented
with the uppermost, red, curve (color online). Escaping L-- and R-- electromagnetic wave energy
peaks at $0.1 \%$ of the beam initial energy. This is comparable to the other mechanisms for the Type--III
radio burst generation.

\section{Discussion and summary of the results \label{SecSummary}}

1.5--dimensional system of the magnetised background Maxwellian plasma and the injected low density relativistic beam
of electrons is numerically studied by utilising PIC code EPOCH. Background magnetic field is constant in each
run and it mimics a magnetic field line from the Sun to the Earth. Background plasma density falls
from the $n_{Sun}=10^{14} m^{-3}$, on the left hand side of the box, to the $n_{Earth}=10^{-6} m^{-6}$ on
the right hand side of the box. Following Ref.\cite{2011PhPl...18e2903T}, the non--gyrotropic beam
injected in such a system is found
to release the freely escaping electromagnetic radiation. 
Ref.\cite{2013PhPl...Schmitz} established the electromagnetic radiation being generated 
due to the $(\omega,k)$--space drift ($i.e.$ wave refraction),
when the non--gyrotropic beam propagates along the decreasing density profile. 
In this study a new feature, the wave component travelling backwards, 
is found in the perpendicular components of the electromagnetic field. 
In order to complete the picture of the non--gyrotropic beam of the Type--III emission mechanism,
the features, characterisation and the exciting mechanism of the backwards propagating waves are
presented in this study.

Constant background magnetic field was varied in different runs, thereby changing the electron--cyclotron frequency.
As a consequence, the frequency of the wave propagating backwards from the front of the beam is found 
to be proportional to the electron--cyclotron frequency. Fast Fourier Transform analysis of the $(\omega, k)$--space is 
performed and the backwards travelling harmonics are identified as the whistler mode waves. In order to infer
possible mechanism for the wave generation, combination of the
Doppler-shifted relativistic resonance condition and the whistler dispersion equation is 
solved. The wave is found to be generated by the normal Doppler-shifted relativistic
resonance. 
Width of the beam has been varied and, as expected from Eq.(\ref{eq:cubic}), resonant excitation of whistlers
was not affected. However, when the beam velocity is varied, the change of the resonance is apparent, in accordance
with the solutions of Eq.(\ref{eq:cubic}). Large fraction, $\approx 58 \%$, 
of the energy of the perpendicular electromagnetic field components goes into whistlers. In turn, the latter 
constitute $0.096 \%$  of the electron beam initial energy.
The perpendicular elecromegnetic energy of the generated electromagnetic radiation at its peak amounts to
$\approx 0.054 \%$ of the initial beam kinetic energy.

A wave component travelling backwards in the numerical system in which non--gyrotropic beam excites
electromagnetic radiation resembling the Type--III emission is conclusively identified as the whistler mode
 wave. It is shown to be excited by the normal Doppler-shifted relativistic resonance.
Whistlers propagating backwards while Type--III propagates
ahead of the electron beam is a possible observable for the in--situ measurements by the spacecrafts in the solar
wind. Simultaneous detection of the related whistlers and the Type--III burst could be an 
observational confirmation of the non--gyrotropic beam mechanism of exciting the Type--III emission taking place.  
To our knowledge only associated Langmuir waves have been detected in--situ \cite{1981ApJ...251..364L}.

{\bf{Acknowledgements}}    \ \
The EPOCH code used in this research was developed under UK Engineering and Physics Sciences Research Council
grants EP/G054940/1, EP/G055165/1 and EP/G056803/1. The simulations were run on the HPC clusters Andromeda of the Astronomy Unit, Queen Mary University of London, and Minerva of the STFC--funded UKMHD consortium at the 
University of St. Andrews and the University of Warwick. M.S. is funded by the Leverhulme Trust 
Research Project Grant RPG--311. D.T. is financially supported by STFC consolidated Grant ST/J001546/1, 
the Leverhulme Trust Research Project Grant RPG--311 and HEFCE--funded South East Physics Network (SEPNET). 
Authors would like to thank the anonymous referee whose suggestions helped in improving the article. 

\end{document}